\documentclass[twocolumn,showpacs,superscriptaddress,prc]{revtex4}%
\usepackage{amsfonts}
\usepackage{amsmath}
\usepackage{amssymb}
\usepackage{graphicx}%
\begin{document}
\preprint{JLAB-THY-06-xxx}
\preprint{KSUCNR-206-06}
\preprint{nucl-th/06mmnnn}
\title{\hspace*{\fill}\texttt{{\normalsize JLAB-THY-06-497, KSUCNR-206-06}}\\[1ex] Quark-gluon vertex dressing and meson masses\footnotetext{Notice: This
manuscript has been authored by The Southeastern Universities Research
Association, Inc. under Contract No. DE-AC05-84150 with the U.S. Department of
Energy. The United States Government retains and the publisher, by accepting
the article for publication, acknowledges that the United States Government
retains a non-exclusive, paid-up, irrevocable, world wide license to publish
or reproduce the published form of this manuscript, or allow others to do so,
for United States Government purposes.}\\beyond ladder-rainbow truncation.}
\author{Hrayr H. Matevosyan}
\affiliation{Louisiana State University, Department of Physics \&
              Astronomy, 202 Nicholson Hall, Tower Dr., LA 70803, USA}
 \affiliation{Thomas Jefferson National Accelerator Facility, 12000
              Jefferson Ave., Newport News, VA 23606, USA}
 \affiliation{Laboratory of Theoretical Physics, JINR, Dubna, Russian
              Federation}
\author{Anthony W. Thomas}
\affiliation{Thomas Jefferson National Accelerator Facility, 12000 Jefferson Ave., Newport
News, VA 23606, USA}
\author{Peter C. Tandy}
\affiliation{Center for Nuclear Research, Department of Physics, Kent State University,
Kent, Ohio 44242, USA}

\begin{abstract}
We include a generalized infinite class of quark-gluon vertex dressing
diagrams in a study of how dynamics beyond the ladder-rainbow truncation
influences the Bethe-Salpeter description of light quark pseudoscalar and
vector mesons. \ \ \ The diagrammatic specification of the vertex is mapped
into a corresponding specification of the Bethe-Salpeter kernel, which preserves
chiral symmetry. \ \ \ This study adopts the algebraic format afforded by the
simple interaction kernel used in previous work on this topic. \ \ The new
feature of the present work is that in every diagram summed for the vertex and
the corresponding Bethe-Salpeter kernel, each quark-gluon vertex is required
to be the self-consistent vertex solution. \ \ We also adopt from previous
work the effective accounting for the role of the explicitly non-Abelian three
gluon coupling in a global manner through one parameter determined from recent
lattice-QCD data for the vertex. \ \ \ With the more consistent vertex used
here, the error in ladder-rainbow truncation for vector mesons is never more
than 10\% as the current quark mass is varied from the u/d region to the b region.

\end{abstract}
\date{\today}
\maketitle

\section{Introduction}

In recent years there has been significant progress in the study of the
spectrum of hadrons, and their non-perturbative structure and form factors,
through approaches that are manifestly covariant and which accommodate both
dynamical chiral symmetry breaking (DCSB) and quark
confinement~\cite{Maris:2003vk}. Covariance provides efficient and unambiguous
access to form factors~\cite{Volmer:2000ek,Maris:2000sk,Alkofer:2004yf}.
Consistency with chiral symmetry and its spontaneous breaking is obviously
crucial to prevent the pseudoscalars from artificially influencing the
difficult task of describing and modeling the infrared dynamics; this is a
role better left to other hadronic states that are not so dominated by chiral
symmetry. The associated concept of a constituent quark mass is important and
it is often implemented in models as a constant mass appearing in the
propagator; however this idealization runs into trouble for higher lying
states where the sum of the constituent masses is below the hadron mass. This
difficulty is marginally evident with the $\rho$, but it is inescapable by the
time one has reached the ground state axial vector mesons (e.g., $a_{1}$,
$b_{1}$ mesons)~\cite{Jarecke:2002xd}.

In reality, solutions of the QCD equation of motion for the quark propagator
(quark Dyson-Schwinger equation (DSE)) give a momentum-dependent quark mass
function.   Model calculations, mostly in Landau gauge, typically yield a mass
function that evolves from the current mass value at ultra-violet spacelike
momenta to a value some 0.4~GeV larger in the deep
infrared~\cite{Maris:1997tm}.  The propagator is a gauge-dependent object and 
the gauge dependence of this phenomenon has not been fully explored.  In the
chiral limit, such an enhancement is DCSB.   At finite current mass, models also 
strongly suggest that the enhancement is the same mechanism as DCSB which
has an important influence over the low-lying hadron spectrum.   In the chiral
limit, the scalar term of the quark self-energy, which shows most of the momentum
dependence, plays a dual role as the dominant invariant amplitude of the chiral
pion Bethe-Salpeter equation (BSE) amplitude at low
momenta~\cite{Maris:1998hd}. 
In any process where the spatial extent of the pion plays an important role,  
the running of the
quark mass function is likewise crucial to an efficient symmetry-preserving
description.  Otherwise a theoretical model is fighting symmetries.   An example
is provided by the pion charge form factor above the chiral symmetry breaking
scale, i.e., \mbox{$Q^2 > m_\rho$}.   It is this
large value of the dressed quark mass function at low spacelike momentum that
leads, in model solutions of the quark DSE, to $|p^{2}|\neq M^2(p^{2})$ within a
significant domain of timelike momenta where these models can be trusted. 
For example, this is sufficient to prevent spurious $q\bar{q}$ production 
thresholds in light quark hadrons
below about 2~GeV~\cite{Jarecke:2002xd}.

The task of maintaining manifest covariance, DCSB, a running quark mass
function and an explicit substructure in terms of confined quarks is often met
by models defined as truncations of the DSEs of
QCD~\cite{Roberts:2000aa,Alkofer:2000wg,Maris:2003vk}. For practical reasons
the equations must be truncated to decouple arbitrarily high order n-point
functions from the set of low order n-point functions used to construct
observables. A common truncation scheme is the rainbow-ladder truncation. Here
the one-loop gluon dressing of the quark (with bare gluon-quark vertices) is
used self-consistently to generate the quark propagator. In general the kernel
$K$ of the Bethe-Salpeter equation is given in terms of the quark self-energy
$\Sigma$ by a functional relation dictated by chiral
symmetry~\cite{Munczek:1995zz}. This preserves the Ward-Takahashi identity for
the color singlet axial vector vertex and ensures that chiral pseudoscalars
will remain massless, independent of model details. With a rainbow
self-energy, this relation yields the ladder BSE kernel. To go beyond this
level, one needs to realize that the exact quark self-energy is given by the
same structure except that one of the gluon-quark vertices is fully dressed.
It is the vertex dressing that generates the terms in $K$ beyond ladder level.
This is the topic we are concerned with in this paper.

The ladder BSE for meson bound states is an integral equation with a one-loop
kernel structure that must allow for the spinor structure of propagators and
the meson amplitudes. With the four-dimensional space-time that one must use
to maintain manifest covariance, and with dynamically generated quark
propagators that one must use to preserve the Ward-Takahashi identities of
chiral symmetry, the numerical task is large. Any scheme for corrections to
the ladder truncation will in general add the complexity of multiple loop
Feynman diagrams involving amplitudes that are only known after solution. For
practical reasons the studies that have been able to investigate hadron states
beyond ladder-rainbow (LR) truncation in recent
years~\cite{Bender:1996bb,Bender:2002as,Watson:2004kd,Bhagwat:2004hn} have
exploited the simplifications following from use of the Munczek-Nemirovsky
(MN) model~\cite{Munczek:1983dx}. In this case the basic element is a delta
function that restricts the exchanged (or gluon) momentum to zero; it reduces 
both the quark DSE and the meson BSE to algebraic equations. There is only one 
parameter--- a strength set by $m_{\rho}$. 

This simplified kernel has no support in the ultra-violet and one must be wary 
of its use for related physics.  Bound state masses are relatively safe in this 
regard, even heavy quark states that sample short distance or large momenta due
to the large quark mass scale present.   Even with the MN model, the DSE solutions 
for the quark propagators have the correct power law behavior, and support connected
to the current quark mass, in the ultra-violet, apart from log corrections.   
The dominant qualitative features of DSE solutions of realistic model are preserved 
in the MN model: large infrared strength giving DCSB and the (confining) absence 
of a mass pole.   Our analysis is not aimed to provide a serious representation 
of experimental data; rather we aim at some understanding, even if it is quite 
qualitative, of the relative importance of classes of higher-order diagrams for 
the BSE kernel for bound states.   Because of the inherent complexity brought by 
use of a momentum distribution as a kernel, there is little information available 
in the literature on this topic.  To obtain such information, we feel the price 
paid by dispensing with a clear connection to perturbative QCD is worthwhile
in the initial stages. 

The first study of the correction to ladder-rainbow truncation was made in
this context in Ref.~\cite{Bender:1996bb} where a one-gluon exchange dressing
of the quark-gluon vertex was implemented for pseudoscalar and vector mesons
and scalar and axial vector diquark correlations.
Subsequently~\cite{Bender:2002as} it was realized that the algebraic structure
allowed a recursive implementation of the ladder series of diagrams for the
quark-gluon vertex as well as an implementation of the corresponding series of
diagrams for the chiral symmetry-preserving BSE kernel. So far as we are aware
this was the first solution of a BSE equation in which the kernel contained
the effects from an infinite number of loops. In these works the chiral
pseudoscalars remained massless independent of the model parameter, $m_{\rho}$
received corrections of order 10\% from ladder dressing of the vertex, and the
diquark states evident at ladder-rainbow level were removed from the spectrum
by the dressing effects. The influence of vertex dressing upon the quark
propagator was also studied.

There is very little in the way of guidance from realistic non-perturbative
non-Abelian models of the infrared structure of the quark-gluon vertex. It has
often been assumed, e.g., see Ref.~\cite{Fischer:2003rp}, that a reasonable
beginning is the Ball-Chiu~\cite{Ball:1980ay} or
Curtis-Pennington~\cite{Curtis:1990zs} Abelian Ansatz times the appropriate
color matrix. These Abelian descriptions of the momentum dependence satisfy
the Abelian vector Ward-Takahashi identity, and their use makes the implicit
assumption that this might be a good enough approximation to the corresponding
identity for QCD, namely the Slavnov-Taylor identity for the color octet
vertex~\cite{Marciano:1977su}. The use of an explicit ladder sum for the gluon
vertex provides easy access to the chiral symmetry preserving BSE kernel and
receives some motivation from the fact that a ladder-summed photon-fermion
vertex combines with the rainbow approximation for the fermion propagator to
preserve the Ward-Takahashi identity for that vertex.

However when initial results from lattice-QCD simulations of the gluon-quark
vertex became available~\cite{Skullerud:2003qu,Skullerud:2004gp}, it was
realized~\cite{Bhagwat:2004kj} that the color algebra generated by any ladder
sum for this vertex gives a magnitude and strength for the dominant amplitude
at zero gluon momentum that is qualitatively and quantitatively incompatible
with the lattice data and incompatible with the leading ultra-violet behavior
of the one-loop QCD Slavnov-Taylor identity. The infrared vertex model
developed in Ref.~\cite{Bhagwat:2004kj} made an extension of the fact that the
one-loop QCD color structure introduced by the three-gluon coupling repairs
the deficiency of a purely ladder structure. The color structure of the ladder
class of diagrams produces a weak repulsive vertex, while the color structure
of the three-gluon coupling contribution produces an attractive contribution
that is enhanced by a factor of -$N_{c}^{2}$ at the purely one-loop level.

These observations from Ref.~\cite{Bhagwat:2004kj} were blended with the
algebraic features afforded by the MN model to re-examine the relation between
vertex dressing, the chiral symmetry-preserving BSE kernel, and the resulting
meson spectrum and diquark correlations~\cite{Bhagwat:2004hn}. This approach
introduced one extra parameter (besides the gluon 2-point function strength
and the quark current mass)---an effective net color factor fitted to
lattice-QCD data on the gluon-quark vertex. The net attraction in the vertex,
driven by the explicitly non-Abelian 3-gluon coupling, had a marked effect:
the ladder-rainbow truncation made $m_{\rho}$ 30\% too high compared to the
solution from the completely summed vertex. In other words, the attraction
produced by summed vertex dressing in a non-Abelian context is more important
than previously thought. However in that approach, the structure of the vertex
is such that the coupling of any internal gluon line to a quark, is itself
bare. This is not self-consistent and one can question what effect this
omitted infinite sub-class of vertex dressing and BSE kernel contributions may
have upon the hadron spectrum.

In the present work we extend the analysis of Ref.~\cite{Bhagwat:2004hn} by
the incorporation of a wider class of vertex dressing diagrams. We allow the
coupling of any internal gluon line to a quark to be described by the dressed
vertex at an order consistent with a given total order in the final vertex. In
the limit of the vertex summed to all orders, this becomes the use of the
self-consistent quark-gluon vertex at every internal location in a diagram. We
borrow from previous work the use of the MN model of the 2-point gluon
function to generate an algebraic structure and we again incorporate the
important non-Abelian three-gluon coupling through the device of an effective
net color factor refitted to the lattice data for the vertex. We use the
infinite series of diagrams for the BSE kernel generated from the chiral
symmetry-preserving relation to the quark self-energy. We investigate the
resulting spectrum of pseudoscalar and vector mesons.

In Section II we describe general properties of the quark-gluon vertex and the
relationship with the associated BSE kernel that preserves chiral symmetry.
Information from the Slavnov-Taylor identity for the gluon-quark vertex and
the Ward-Takahashi identity for the color singlet axial-vector vertex is
summarized for relevance to present considerations. We discuss diagrammatic
summations that have been used previously to model the gluon vertex and the
generalized class of diagram considered here. In Section III we introduce the
interaction model that allows an algebraic analysis and we present consequent
results for the gluon-quark vertex and the self-consistent dressed quark
propagator. The associated symmetry-preserving BSE kernel is presented also.
Section IV contains a presentation and analysis of the methods and results for
the meson masses. In Section V there is a summary.

\section{The quark-gluon vertex and the Bethe-Salpeter kernel}

We employ Landau gauge and a Euclidean metric, with: $\{\gamma_{\mu}%
,\gamma_{\nu}\}=2\delta_{\mu\nu}$; $\gamma_{\mu}^{\dagger}=\gamma_{\mu}$; and
$a\cdot b=\sum_{i=1}^{4}a_{i}b_{i}$. The dressed quark-gluon vertex for gluon
momentum $k$ and quark momentum $p$ can be written $ig\,t^{c}\,\Gamma_{\sigma
}(p+k,p)$, where $t^{c}=\lambda^{c}/2$ and $\lambda^{c}$ is an SU(3) color
matrix. In general, $\Gamma_{\sigma}(p+k,p)$ has 12 independent invariant
amplitudes. We are particularly concerned in this work with the vertex at
$k=0$, in which case the general form is%
\begin{multline}
\Gamma_{\sigma}(p)=\alpha_{1}(p^{2})\gamma_{\sigma}+\alpha_{2}(p^{2}%
)\gamma\cdot p~p_{\sigma}-\alpha_{3}(p^{2})ip_{\sigma}\label{vertex_form}\\
+\alpha_{4}(p^{2})i\gamma_{\sigma}~\gamma\cdot p
\end{multline}
where $\alpha_{i}(p^{2})$ are invariant amplitudes. In the model studies of
Refs.~\cite{Bender:2002as} and \cite{Bhagwat:2004hn} that we build upon, one
finds $\alpha_{4}=0$; this will also be the case here.

As we shall discuss later, we wish to utilize the functional relation that
enables the BSE kernel to be generated from the quark self-energy so that
chiral symmetry is preserved. This requires the vertex to be represented in
terms of a set of explicit Feynman diagrams. Some exact results are known for
the vertex at 1-loop order in QCD~\cite{Davydychev:2000rt}. In Landau gauge
and to $\mathcal{O}(g^{2})$, i.e., to 1-loop, the amplitude $\Gamma_{\sigma}$
is given by
\begin{equation}
\Gamma_{\sigma}^{(1)}(p+k,p)=Z_{1}\gamma_{\sigma}+\Gamma_{\sigma}^{\mathrm{A}%
}(p+k,p)+\Gamma_{\sigma}^{\mathrm{NA}}(p+k,p), \label{vert1loop}%
\end{equation}
with
\begin{multline}
\Gamma_{\sigma}^{\mathrm{A}}(p+k,p)=-({C_{\mathrm{F}}}-\frac{C_{\mathrm{A}}%
}{2})\int_{q}^{\Lambda}g^{2}D_{\mu\nu}(p-q)\gamma_{\mu}\label{VertA}\\
\times S_{0}(q+k)\gamma_{\sigma}S_{0}(q)\gamma_{\nu},
\end{multline}
and%
\begin{multline}
\Gamma_{\sigma}^{\mathrm{NA}}(p+k,p)=-\frac{C_{\mathrm{A}}}{2}\int
_{q}^{\Lambda}g^{2}\gamma_{\mu}S_{0}(p-q)\gamma_{\nu}D_{\mu\mu^{\prime}%
}(q+k)\label{VertNA}\\
\times i\Gamma_{\mu^{\prime}\nu^{\prime}\sigma}^{3g}(q+k,q)D_{\nu^{\prime}\nu
}(q),
\end{multline}
where $\int_{q}^{\Lambda}=\int^{\Lambda}d^{4}q/(2\pi)^{4}$ denotes a loop
integral regularized in a translationally-invariant manner at mass-scale
$\Lambda$. Here $Z_{1}(\mu^{2},\Lambda^{2})$ is the vertex renormalization
constant to ensure $\Gamma_{\sigma}=\gamma_{\sigma}$ at renormalization scale
$\mu$. The following quantities are bare: the three-gluon vertex
$ig\,f^{\mathrm{abc}}\,\Gamma_{\mu\nu\sigma}^{3g}(q+k,q)$, the quark
propagator $S_{0}(p)$, and the gluon propagator $D_{\mu\nu}(q)=T_{\mu\nu
}(q)D_{0}(q^{2})$, where $T_{\mu\nu}(q)$ is the transverse projector. The next
order terms in Eq.~(\ref{vert1loop}) are $\mathcal{O}(g^{3})$: the
contribution involving the four-gluon vertex, and $\mathcal{O}(g^{4})$:
contributions from crossed-box and two-rung gluon ladder diagrams, and 1-loop
dressing of the triple-gluon vertex, etc. The color factors in
Eqs.~(\ref{VertA})and (\ref{VertNA}) are given by
\begin{gather}
t^{a}t^{b}t^{a}=({C_{\mathrm{F}}}-\frac{C_{\mathrm{A}}}{2})t^{b}=-\frac
{1}{2N_{c}}t^{b}\nonumber\\
t^{a}f^{abc}t^{b}=\frac{C_{\mathrm{A}}}{2}it^{c}=\frac{N_{c}}{2}%
it^{c}\nonumber\\
t^{a}t^{a}={C_{\mathrm{F}}}\mathbf{1_{c}}=\frac{(N_{c}^{2}-1)}{2N_{c}%
}\mathbf{1_{c}}. \label{colorfacts}%
\end{gather}
In contrast, for the color singlet vector vertex, i.e., for the strong
dressing of the quark-photon vertex, one has the one-loop Abelian result
\begin{multline}
\tilde{\Gamma}_{\sigma}^{(1)}(p+k,p)=\tilde{Z}_{1}\gamma_{\sigma
}-{C_{\mathrm{F}}}\int_{q}^{\Lambda}g^{2}D_{\mu\nu}(p-q)\gamma_{\mu}%
S_{0}(q+k)\label{QED1loop}\\
\times\gamma_{\sigma}\,S_{0}(q)\gamma_{\nu}.
\end{multline}

To motivate the approximate vertex used in the present study, we note that the
local color SU(3) gauge invariance of the QCD action gives the Slavnov-Taylor
identity~\cite{Marciano:1977su} for the gluon vertex
\begin{multline}
\lefteqn{k_{\mu}i\Gamma_{\mu}(p+k,p)=G(k^{2})\left\{  \left[  1-B(p,k)\right]
S(p+k)^{-1}\right.  }\label{STI}\\
\left.  -S(p)^{-1}\left[  1-B(p,k)\right]  \right\}  ,
\end{multline}
which relates the divergence of the vertex to the quark propagator $S(p)$, the
dressing function $G(k^{2})$ of the ghost propagator $-G(k^{2})/k^{2}$, and
the ghost-quark scattering kernel $B(p,k)$, all consistently renormalized.
Even though there is no explicit ghost content evident in the 1-loop vertex
Eq.~(\ref{vert1loop}), it does satisfy this identity at one-loop
order~\cite{Davydychev:2000rt}.

The dressed quark propagator appearing in Eq.~(\ref{STI}) is the solution to
the gap equation, or the quark Dyson-Schwinger equation, which is
\begin{multline}
S^{-1}(p)=Z_{2}\,S_{0}^{-1}(p)+{C_{\mathrm{F}}}\,Z_{1}\int_{q}^{\Lambda}%
g^{2}D_{\mu\nu}(p-q)\,\gamma_{\mu}\label{gap_eqn}\\
\times S(q)\Gamma_{\nu}(q,p),
\end{multline}
where $S_{0}^{-1}(p)=i\gamma\cdot p+m_{bm}$, $m_{bm}$ is the bare current
quark mass, and $Z_{2}(\mu^{2},\Lambda^{2})$ is the quark wave function
renormalization constant. The general form for $S(p)^{-1}$ is
\begin{equation}
S(p)^{-1}=i\gamma\cdot p\,A(p^{2},\mu^{2})+B(p^{2},\mu^{2})
\label{EQ_QUARK_PROPAGATOR}%
\end{equation}
and the renormalization condition at scale $p^{2}=\mu^{2}$ is $S(p)^{-1}%
\rightarrow$ $i\gamma\cdot p+m(\mu)$ where $m(\mu)$ is the renormalized
current quark mass.

Prior to the recent appearance of quenched lattice-QCD
data~\cite{Skullerud:2003qu,Skullerud:2004gp}, there had been little
information available on the infrared structure of the gluon-quark vertex. The
two $\mathcal{O}(g^{2})$ diagrams of Eq.~(\ref{vert1loop}) can not be expected
to be adequate there. A common assumption~\cite{Fischer:2003rp} has been to
adopt an Abelian vertex Ansatz, such as the Ball-Chiu~\cite{Ball:1980ay} or
Curtis-Pennington~\cite{Curtis:1990zs} forms and attach the appropriate color
matrix. In the case of an Abelian U(1) gauge theory, the counterpart to
Eq.~(\ref{STI}) is the Ward-Takahashi identity (WTI)
\begin{equation}
k_{\mu}\,i\tilde{\Gamma}_{\mu}(p+k,p)=S(p+k)^{-1}-S(p)^{-1}. \label{vWTI}%
\end{equation}
At $k=0$, the Abelian vertex $\tilde{\Gamma}_{\mu}$ has the same general form
as given earlier in Eq.~(\ref{vertex_form}). The Ward identity $\tilde{\Gamma
}_{\sigma}(p)=$ $-i\partial S^{-1}(p)/\partial p_{\sigma}$ yields:
$\tilde{\alpha}_{1}=A(p^{2})$, $\tilde{\alpha}_{2}=$ $2\,A^{\prime}(p^{2})$,
and $\tilde{\alpha}_{3}=$ $2\,B^{\prime}(p^{2})$, where $f^{\prime}=$
$\partial f(p^{2})/\partial p^{2}$. However, even if the Abelian Ansatz,
$ig\,t^{c}\,\tilde{\Gamma}_{\sigma}(p)$, were to be adopted for the gluon
vertex, it would not help in the present context because we need a
representation in terms of an explicit set of Feynman diagrams for the
resulting self-energy, in order to determine the symmetry-preserving BSE kernel.

In Ref.~\cite{Bender:2002as} a study was made of a ladder summation Ansatz for
the gluon vertex based on just the Abelian-like gluon exchange diagram of
Eq.~(\ref{VertA}); the symmetry-preserving BSE kernel was generated and used
to explore meson and diquark masses. The vertex was generated by iterative and
recursive techniques and, after convergence, is equivalent to solution of the
integral equation
\begin{multline}
\Gamma_{\sigma}(p+k,p)=Z_{1}\gamma_{\sigma}-({C_{\mathrm{F}}}-\frac
{C_{\mathrm{A}}}{2})\int_{q}^{\Lambda}g^{2}D_{\mu\nu}(p-q)\gamma_{\mu
}\label{detmoldvertex}\\
\times S(q+k)\Gamma_{\sigma}(q+k,q)S(q)\gamma_{\nu}.
\end{multline}
Here, at any order of iteration, the quark propagator is calculated by using
the same vertex in the gap equation, Eq.~(\ref{gap_eqn}). Is this ladder sum a
good approximation to the gluon-quark vertex, particularly in the infrared?
The quenched lattice-QCD data indicates that the answer is no. The lattice
data clearly gives $\alpha_{1}(p^{2})>1$ for all available $p^{2}$ and the
infrared limit appears to be $\alpha_{1}(0)\gtrsim2.2$. The ladder summation
based on Eq.~(\ref{detmoldvertex}) gives $\alpha_{1}(p^{2})<1$, with infrared
limit $\alpha_{1}(0)\approx0.94$. The 1-loop QCD analysis indicates that in
the ultra-violet $\alpha_{1}(p^{2})$ approaches unity from
above~\cite{Davydychev:2000rt}, while the recent model
vertex~\cite{Bhagwat:2004kj}, based on a non-perturbative extension of the two
1-loop diagrams from Eq.~(\ref{vert1loop}), yields $\alpha_{1}(p^{2})>1$ for
all $p^{2}$, and agrees quite well with the lattice data.

The reason for this problem can be seen from the color factors associated with
the two 1-loop diagrams, Eq.~(\ref{VertA}) and Eq.~(\ref{VertNA}), which are
the leading terms in the ultra-violet region. The ladder sum in
Eq.~(\ref{detmoldvertex}) is built on the least significant of the two
diagrams; the color factor of the omitted 3-gluon term is $-N_{c}^{2}$ times
that of the retained term. The relative contribution to the Slavnov-Taylor
identity, Eq.~(\ref{STI}), from that term is of the same order at 1-loop. More
generally, as discussed in Ref.~\cite{Bhagwat:2004hn}, if $G(k^{2}%
)(1-B(p,k))>0$ persists into the non-perturbative region, one can expect
$\alpha_{1}(p^{2})>1$. One can also expect to obtain the wrong sign for
$\alpha_{1}(p^{2})-1$ if a model kernel has the wrong sign. This is the case
with the Abelian-like ladder sum, Eq.~(\ref{detmoldvertex}). Note that in an
Abelian $U(1)$ gauge theory, e.g., the photon-quark vertex, $\tilde{\alpha
}_{1}(p^{2})=A(p^{2})>1$. An Abelian Ansatz for this amplitude of the
gluon-quark vertex might be quite reasonable, but it cannot be simulated by an
explicit ladder sum---the color algebra prevents it. In analogy with the
photon-quark vertex, where $\tilde{\alpha}_{1}(p^{2})>1$ is correlated with
the spectral density being positive definite as the timelike region is
approached, the gluon-quark vertex dressing has been referred to as an
attractive effect in the infrared spacelike region~\cite{Bhagwat:2004hn}. (Of
course, for the gluon vertex there should be no color octet bound states and
no positive spectral density in the timelike region.) The 3-gluon coupling is
a strong source of the attraction at low spacelike $p^{2}$; it is $N_{c}^{2}$
times larger than the small repulsive effect of gluon exchange.

The model for $D_{\mu\nu}$ that we employ in this work, described in Section III,
allows us to focus on zero gluon momentum.   In this case, as discussed and
utilized in Ref.~\cite{Bhagwat:2004kj}, the two pQCD 1-loop diagrams for the vertex, 
Eq.~(\ref{VertA}) and Eq.~(\ref{VertNA}), are both closely related to the
momentum derivative of the corresponding quark self-energy, apart from the
differing color factors.  The resulting dependence upon the single quark momentum
variable is similar for each diagram.  Both are 1-loop integrals projected onto
the same Dirac structures.  We adopt the approach of Ref.~\cite{Bhagwat:2004hn}
to the vertex for our algebraic study;  the approach is defined by taking the 
momentum dependence to be  similar even in the infrared and with dressed propagators.   
Thus we combine the two terms and write Eq.~(\ref{vert1loop}) as 
\begin{multline}
\Gamma_{\sigma}^{(1)}(p+k,p)\approx Z_{1}\gamma_{\sigma}-\mathcal{C}%
{C_{\mathrm{F}}}\int_{q}^{\Lambda}g^{2}D_{\mu\nu}(p-q)\gamma_{\mu}%
S_{0}(q+k)\label{effvertex_1loop}\\
\times\gamma_{\sigma}S_{0}(q)\gamma_{\nu},
\end{multline}
with $\mathcal{C}$ being an effective color factor to be determined by a fit to
lattice-QCD data for the vertex.  If the momentum dependence
of the two combined terms from Eq.~(\ref{vert1loop}) was identical, then we
see that $\mathcal{C}=1$; this is equivalent to the Abelian limit. If one were
to omit the 3-gluon term altogether, as in the iterative study in
Ref.~\cite{Bender:2002as}, then $\mathcal{C}=$ $({C_{\mathrm{F}}}%
-\frac{C_{\mathrm{A}}}{2})\,{C_{\mathrm{F}}}^{-1}$, which for $N_{c}=3$, gives
$\mathcal{C}=-1/8$. One expects that the non-Abelian term is necessary for an
effective model and thus that $0<\mathcal{C}<1$. 

This vertex Ansatz allows us to avoid making a model for the dressed 3-gluon vertex
for which there is little in the way of reliable information.  It is implicitly
hoped that the fit of $\mathcal{C}$ to lattice data will effectively compensate for
deficiencies.  Our aim is not the vertex itself but a study of the relative importance
of classes of diagrams for the BSE kernel for meson masses.  This vertex Ansatz allows
an algebraic approach to the BSE meson masses that is quite illustrative of new
qualitative information.

From Eq.~(\ref{effvertex_1loop}), the non-perturbative summation equivalent to the
integral equation
\begin{multline}
\Gamma_{\sigma}(p+k,p)=Z_{1}\gamma_{\sigma}-\mathcal{C}{C_{\mathrm{F}}}%
\int_{q}^{\Lambda}g^{2}D_{\mu\nu}(p-q)\gamma_{\mu}%
S(q+k)\label{eff_ladd_vertex}\\
\times\Gamma_{\sigma}(q+k,q)S(q)\gamma_{\nu},
\end{multline}
is a natural suggestion. This was studied in Ref.~\cite{Bhagwat:2004hn}, with
$S(q)$ being the self-consistent solution of the quark DSE, Eq.~(\ref{gap_eqn}%
), containing the same dressed vertex. A fit to the lattice-QCD data for the
vertex gave $\mathcal{C}=0.51$, a value that confirms that attraction by a
mechanism outside the scope of iterated gluon exchange is present.

\begin{figure}
[h]
\begin{center}
\includegraphics[
height=1.446in,
width=2.0349in
]%
{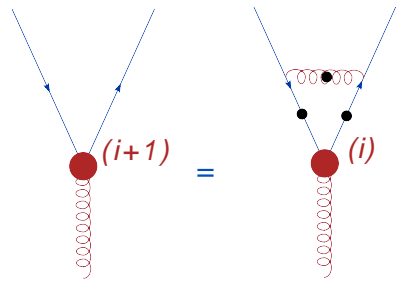}%
\caption{(Color Online) The iterative relation for successive terms in the
ladder-summed vertex. Here the large filled circles denote the dressed
quark-gluon vertex, the numbers in the parenthesis denote the
numbers of gluon lines contributing to the particular vertices and the small
filled circles denote that the propagators are fully
dressed. Note that an important non-Abelian term is approximately accounted for
by the effective color factor $\mathcal{C}$ as described in the text.}%
\label{PLOT_LADDER_VERTEX}%
\end{center}
\end{figure}

An iterative representation is useful: $\Gamma_{\mu}=\sum_{i=0}\,\Gamma_{\mu
}^{i}$, where $\Gamma_{\mu}^{0}=Z_{1}\,\gamma_{\mu}$, and $i$ labels the
number of internal gluon lines. The contribution with $i+1$ internal gluon
lines is obtained from the $i^{th}$ contribution by adding one gluon ladder.
This is schematically depicted in Fig.~\ref{PLOT_LADDER_VERTEX}.%

\subsection{A wider class of quark-gluon vertex dressing}

\begin{figure}
[h]
\begin{center}
\includegraphics[
height=1.318in,
width=2.0349in
]%
{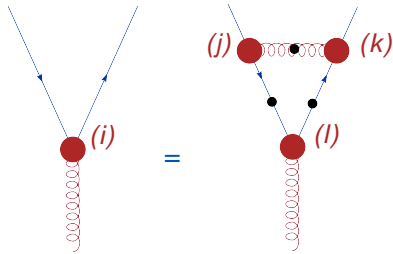}%
\caption{(Color Online) The iterative relation for the enlarged class of
dressing diagrams considered in this work. Here the large filled circles denote
the dressed quark-gluon vertex, the numbers in the parenthesis
denote the numbers of gluon lines contributing to the particular vertices
(with $j+k+l+1\equiv i$) and the small filled circles denote
that the propagators are fully dressed. The vertex contribution with $i$
internal gluon lines is obtained from vertex contributions with less gluon
lines. Note that an important effect of the non-Abelian 3-gluon coupling is
approximately accounted for by the effective color factor $\mathcal{C}$ as
described in the text.}%
\label{PLOT_IMPROVED_VERTEX}%
\end{center}
\end{figure}

The enlarged class of dressing diagrams considered in this work is obtained
iteratively as depicted in Fig.~{\ref{PLOT_IMPROVED_VERTEX}} The contribution
with $i$ internal gluon lines is generated from three contributions having a
smaller number of gluon lines by adding one gluon ladder with dressed
vertices. If the number of gluon lines in the three vertex contributions are
denoted $j,k$ and $l$, then summation is made over $j,k$ and $l$ such that
$j+k+l+1=i$. Again, $\Gamma_{\mu}=\sum_{i=0}\,\Gamma_{\mu}^{i}$. The iterative
scheme is described by
\begin{multline}
\Gamma_{\mu}^{i}(p+k,p)=-\mathcal{C}{C_{\mathrm{F}}}\sum
_{\substack{j,k,l\\i=j+k+l+1}}\int_{q}^{\Lambda}g^{2}D_{\sigma\nu
}(p-q)\label{EQ_IMPROVED_VERTEX_ITERATION}\\
\times\Gamma_{\sigma}^{j}(p+k,q+k)S(q+k)\Gamma_{\mu}^{l}(q+k,q)
S(q)\Gamma_{\nu}^{k}(q,p),
\end{multline}
for $i\geq1$.

If the iteration is carried to all orders, the equivalent integral equation
is
\begin{multline}
\Gamma_{\mu}(p+k,p)=Z_{1}\gamma_{\mu}-\mathcal{C}%
{C_{\mathrm{F}}}\int_{q}^{\Lambda}g^{2}D_{\sigma\nu}(p-q)\Gamma_{\sigma}%
(p+k,q+k)\label{EQ_IMPROVED_VERTEX_DSE_RAW}\\
\times S(q+k)\Gamma_{\mu}(q+k,q)S(q)\Gamma_{\nu}(q,p).
\end{multline}
If the iteration is stopped to produce all vertex functions\ with up to $n$
internal two point gluon lines, our improved scheme takes into account
$1+n(n+1)(n+2)/6$ diagrams; the corresponding ladder-summed vertex at that
order contains a subset of $(n+1)$ of these diagrams.%

\begin{figure}
[ptb]
\begin{center}
\includegraphics[
height=1.3041in,
width=3.4437in
]%
{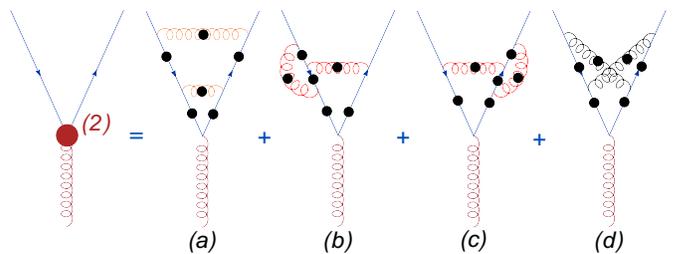}%
\caption{(Color Online) Vertex skeleton diagrams at $\mathcal{O}(g^{5})$. Here the
large filled circle denotes the quark-gluon vertex function dressed to the 
order of two effective gluon kernel lines.    The small filled
circles denote that the
propagators are fully dressed. Previous work included the ladder structure
typified by part (a). The enlarged class of dressing diagrams implemented in
this work includes parts (b) and (c) as well. Non-planar diagrams such as part
(d) are not accommodated by the present approach. We use an effective color
factor to accommodate a major non-Abelian effect from the 3-gluon coupling as
described in the text.}%
\label{PLOT_FULL_VERTEX_2ND_ORDER}%
\end{center}
\end{figure}

In Fig.~{\ref{PLOT_FULL_VERTEX_2ND_ORDER}} we use low order diagrams to
illustrate the more general class of dressing terms included this way. Note
that the included diagrams are restricted to planar diagrams. The contribution
from crossed gluon lines in Fig.~{\ref{PLOT_FULL_VERTEX_2ND_ORDER}}d is not
included. All diagrams of the ladder sum used in Ref.~\cite{Bhagwat:2004hn},
such as Fig.~{\ref{PLOT_FULL_VERTEX_2ND_ORDER}}a, are included; the new
element here is the self-consistent dressing of the internal vertices
illustrated by Fig.~{\ref{PLOT_FULL_VERTEX_2ND_ORDER}}b and
Fig.~{\ref{PLOT_FULL_VERTEX_2ND_ORDER}}c.%

\subsection{Symmetry-preserving Bethe-Salpeter kernel}

The renormalized homogeneous Bethe-Salpeter equation (BSE) for the
quark-antiquark channel, denoted by $M$, can be compactly expressed as%
\begin{equation}
\left[  \Gamma_{M}(k;P)\right]  _{EF}=\int_{q}^{\Lambda}\left[
K(k,q;P)\right]  _{EF}^{GH}\left[  \chi_{M}(q;P)\right]  _{GH},
\label{EQ_BSE_MESON}%
\end{equation}
where $\Gamma_{M}(k;P)$ is the meson Bethe-Salpeter amplitude (BSA), $k$ is
the relative momentum of the quark-antiquark pair and $P$ is their total
momentum; E,...,H represent color, flavor and spinor indices and the BS
wavefunction is
\begin{equation}
\chi_{M}(k;P)=S(k_{+})\Gamma_{M}(k;P)S(k_{-}), \label{EQ_CHI_M}%
\end{equation}
where $k_{\pm}=k\pm\frac{P}{2}$, and $\ K$ is the amputated quark-antiquark
scattering kernel. In general the kernel $K$ is given in terms of the quark
self-energy, $\Sigma$, by a functional relation dictated by chiral
symmetry~\cite{Munczek:1995zz}. This preserves the Ward-Takahashi identity for
the color singlet axial vector vertex and ensures that chiral pseudoscalars
will remain massless, independent of model details.

In a flavor non-singlet channel, and with equal mass quarks, the axial-vector
Ward-Takahashi identity is
\begin{multline}
-iP_{\mu}\Gamma_{\mu}^{5}(p+P,p)=S^{-1}(p+P)\gamma_{5}+\gamma_{5}%
S^{-1}(p)\label{AV_WTI}\\
-2m(\mu)\Gamma^{5}(p+P,p),
\end{multline}
where we have factored out the explicit flavor matrix. The color-singlet
quantities $\Gamma_{\mu}^{5}$ and $\Gamma^{5}$ are the axial-vector vertex and
the pseudoscalar vertex, respectively, and $P$ is the total momentum. The
amplitude $\Gamma_{\mu}^{5}(p+P,p)$ has a pseudoscalar meson pole. A
consequence is that the meson BSE (\ref{EQ_BSE_MESON}) for the (dominant)
$\gamma_{5}$ amplitude at $P^{2}=0$ is equivalent to the chiral limit quark
DSE for $B(p^{2})$ and a non-zero value for the latter (DCSB) necessarily
produces a massless pseudoscalar bound state~\cite{Maris:1998hd}.

The general relation between the BSE kernel, $K$, and the quark self-energy,
$\Sigma$, can be expressed through the functional
derivative~\cite{Munczek:1995zz}
\begin{equation}
K(x^{\prime},y^{\prime};x,y)=-\frac{\delta}{\delta S(x,y)}\Sigma(x^{\prime
},y^{\prime}). \label{Kernelderiv}%
\end{equation}
It is to be understood that this procedure is defined in the presence of a
bilocal external source for $\bar{q}q$ and thus $S$ and $\Sigma$ are not
translationally invariant until the source is set to zero after the
differentiation. An appropriate formulation is the Cornwall-Jackiw-Tomboulis
effective action~\cite{Cornwall:1974vz}. In this context, the above coordinate
space formulation ensures the correct number of independent space-time
variables will be manifest. Fourier transformation of that 4-point function to
momentum representation produces $K(p,q;P)$ having the correct momentum flow
appropriate to the BSE kernel for total momentum $P$.

The constructive scheme of Ref.~\cite{Bender:1996bb} is an example of this
relation as applied order by order to a Feynman diagram expansion for
$\Sigma(p)$. An internal quark propagator $S(q)$ is removed and the momentum
flow is adjusted to account for injection of momentum $P$ at that point. The
number of such contributions coming from one self-energy diagram is the number
of internal quark propagators. Hence the rainbow self-energy generates the
ladder BSE kernel. A 2-loop self-energy diagram (i.e., from 1-loop vertex
dressing) generates 3 terms for the BSE kernel. One can confirm that the
axial-vector Ward-Takahashi identity is preserved. Similarly, the vector
Ward-Takahashi identity is also preserved.

To be more specific, with the discrete indices made explicit, we apply
\begin{equation}
K_{EF}^{GH}=-\frac{\delta\Sigma_{EF}}{\delta S_{GH}},\label{EQ_KERNEL_DER}%
\end{equation}
to the self-energy given by the second term on the RHS of Eq.~(\ref{gap_eqn}).
After a decomposition
\begin{equation}
\Sigma(k)=\sum_{n=0}^{\infty}\Sigma^{n}(k),\label{EQ_SELF_ENERGY_DECOMP}%
\end{equation}
according to the number $n$ of gluon kernels in the vertex defined by
\begin{equation}
\Sigma^{n}(k)={C_{\mathrm{F}}}\int_{q}^{\Lambda}g^{2}D_{\mu\nu}(k-q)\gamma
_{\mu}\,S(q)\Gamma_{\nu}^{n}(q,k),\label{EQ_SELF_ENERGY}%
\end{equation}
for $n\geq1$, with%
\begin{equation}
\Sigma^{0}(k)=m_{bm}+{C_{\mathrm{F}}}\int_{q}^{\Lambda}g^{2}D_{\mu\nu
}(k-q)\gamma_{\mu}S(q)\gamma_{\nu},\label{EQ_SELF_ENERGY_0}%
\end{equation}

The order $n$ contribution to the BSE kernel is
\begin{multline}
\left[  K^{n}(k,q;P)\right]  _{EF}^{GH}=\label{EQ_KERNEL_N_LOOP}\\
-{C_{\mathrm{F}}}g^{2}D_{\mu\nu}(k-q)\left[  \gamma_{\mu}\right]  _{EG}\left[
\Gamma_{\nu}^{n}(q_{-},k_{-})\right]  _{HF}\\
-{C_{\mathrm{F}}}\int_{l}^{\Lambda}g^{2}D_{\mu\nu}(k-l)\left[  \gamma_{\mu
}S(l_{+})\right]  _{EL}\\
\times\frac{\delta}{\delta S_{GH}(q_{\pm})}\left[  \Gamma_{\nu}^{n}%
(l_{-},k_{-})\right]  _{LF}.
\end{multline}

This format is the same as used in Refs.~\cite{Bender:2002as} and
\cite{Bhagwat:2004hn}, except that here the content of $\Gamma_{\nu}^{n}$ is
more extensive. With a bare vertex the first term of
Eq.~(\ref{EQ_KERNEL_N_LOOP}) produces the ladder kernel and the second term is
zero. With a vertex up to 1-loop ($n=1$), the first term of
Eq.~(\ref{EQ_KERNEL_N_LOOP}) produces the ladder term plus a 1-loop correction
to one vertex; the second term produces two terms: a 1-loop correction to the
other vertex and a non-planar term corresponding to crossed gluon lines. These
three corrections to the ladder kernel have the same structure as the kernels shown in
parts (b), (c), and (d) of Fig.~{\ref{PLOT_FULL_VERTEX_2ND_ORDER}}. At higher
order, $n>1$, the BSE kernel produced in the present work departs from that
considered in Ref.~\cite{Bhagwat:2004hn}.

After substitution of Eq.~(\ref{EQ_KERNEL_N_LOOP}) into the BSE
Eq.~(\ref{EQ_BSE_MESON}), and with a change of variables, the meson BSE
becomes
\begin{multline}
\Gamma_{M}(k;P)=-{C_{\mathrm{F}}}\int_{q}^{\Lambda}g^{2}D_{\mu\nu}%
(k-q)\gamma_{\mu}\label{EQ_BSE_MODEL}\\
\times\left[  \chi_{M}(q;P)\Gamma_{\nu}(q_{-},k_{-})+S(q_{+})\Lambda_{M\nu
}(q,k;P)\right]  ,
\end{multline}
where we denote by $\Lambda_{M\nu}$ the summation to all orders of the
functional derivative of the vertex as indicated in
Eq.~(\ref{EQ_KERNEL_N_LOOP}). In particular,
\begin{equation}
\Lambda_{M\nu}(q,k;P)=\sum_{n=0}^{\infty}\Lambda_{M\nu}^{n}(q,k;P),
\label{EQ_LAMBDA_DECOMP}%
\end{equation}
with%
\begin{multline}
\left[  \Lambda_{M\nu}^{n}(q,k;P)\right]  _{LF}=\int_{l}^{\Lambda}\frac
{\delta}{\delta S_{GH}(l_{\pm})}\left[  \Gamma_{\nu}^{n}(q_{-},k_{-})\right]
_{LF}\label{EQ_LAMBDA_N_ORDER}\\
\times\left[  \chi_{M}(l;P)\right]  _{GH}.
\end{multline}

The vertex iteration given in Eq.~(\ref{EQ_IMPROVED_VERTEX_ITERATION})
produces the recurrence formula for the $\Lambda_{M\nu}^{n}$%
\begin{gather}
\Lambda_{M\nu}^{n}(q,k;P)=-\mathcal{C}{C_{\mathrm{F}}}\sum
_{\substack{j,k,h\\n=j+k+h+1}}\nonumber\\
\left.
\begin{array}
[c]{c}%
\left[  \int_{t}^{\Lambda}g^{2}D_{\rho\sigma}(q-t)\Gamma_{\rho}^{j}%
(q_{+},t_{+})\chi_{M}(t;P)\Gamma_{\nu}^{k}(t_{-},t_{-}+k-q)\right.  \\
\times S(t_{-}+k-q)\Gamma_{\sigma}^{h}(t_{-}+k-q,k_{-})
\end{array}
\right.  \nonumber\\
\left.
\begin{array}
[c]{c}%
+\int_{t}^{\Lambda}g^{2}D_{\rho\sigma}(k-t)\Gamma_{\rho}^{j}(q_{+}%
,t_{+}+q-k)S(t_{+}+q-k)\\
\times\Gamma_{\nu}^{k}(t_{+}+q-k,t_{+})\chi_{M}(t;P)\Gamma_{\sigma}^{h}%
(t_{-},k_{-})
\end{array}
\right.  \nonumber\\
\left.
\begin{array}
[c]{c}%
+\int_{t}^{\Lambda}g^{2}D_{\rho\sigma}(q-t)\Lambda_{M\rho}^{j}(q,t;P)S(t_{-}%
)\Gamma_{\nu}^{k}(t_{-},t_{-}+k-q)\\
\times S(t_{-}+k-q)\Gamma_{\sigma}^{h}(t_{-}+k-q,k_{-})
\end{array}
\right.  \nonumber\\
\left.
\begin{array}
[c]{c}%
+\int_{t}^{\Lambda}g^{2}D_{\rho\sigma}(q-t)\Gamma_{\rho}^{j}(q_{+}%
,t_{+})S(t_{+})\Lambda_{M\nu}^{k}(t,t+k-q;P)\\
\times S(t_{-}+k-q)\Gamma_{\sigma}^{h}(t_{-}+k-q,k_{-})
\end{array}
\right.  \nonumber\\
\left.
\begin{array}
[c]{c}%
+\int_{t}^{\Lambda}g^{2}D_{\rho\sigma}(q-t)\Gamma_{\rho}^{j}(q_{+}%
,t_{+})S(t_{+})\Gamma_{\nu}^{k}(t_{+},t_{+}+k-q)\\
\left.  \times S(t_{+}+k-q)\Lambda_{M\sigma}^{h}(t+k-q,k;P)\right]
\end{array}
\right.  \label{EQ_LAMBDA_RECUR}%
\end{gather}
where $\Lambda_{M\nu}^{0}(q,k;P)=0$.%
\begin{figure}
[ptb]
\begin{center}
\includegraphics[
height=0.723in,
width=3.4126in
]%
{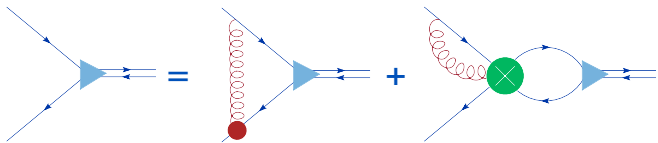}%
\caption{(Color Online) Kernel decomposition. The filled triangles represent the
meson BSAs, the filled circle represents the dressed quark-gluon vertex 
and the crossed circle represents the $\Lambda$ function.}%
\label{PLOT_KERNEL_DECOMP}%
\end{center}
\end{figure}
\begin{figure}
[ptbptb]
\begin{center}
\includegraphics[
height=1.606in,
width=3.4126in
]%
{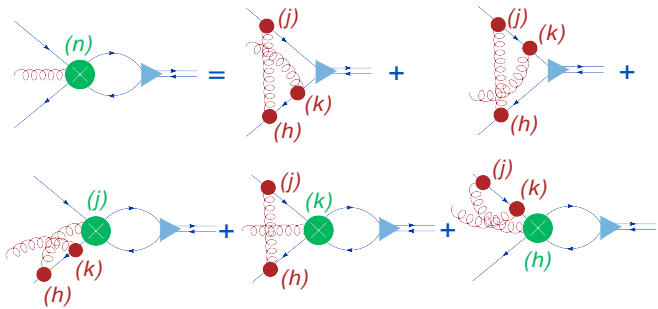}%
\caption{(Color Online) $\Lambda$ function decomposition. The filled triangles
represent the meson BSAs, the filled circles represent the dressed quark-gluon
vertices and the crossed circles represent the $\Lambda$
functions. The numbers in the parenthesis denote the numbers of gluon lines
contributing to the particular functions.}%
\label{PLOT_LAMBDA_DECOMP}%
\end{center}
\end{figure}

The structure of the $q\bar{q}$ BS kernel produced by Eq.\ (\ref{EQ_BSE_MODEL}%
) and Eq. (\ref{EQ_LAMBDA_RECUR}) is schematically depicted in Figs.
\ref{PLOT_KERNEL_DECOMP} and \ref{PLOT_LAMBDA_DECOMP}. With a general
interaction kernel, $g^{2}D_{\rho\sigma}$, it is exceedingly difficult to
implement this formal recurrence relation to obtain a BS kernel because of
overlapping multiple integrals that compound rapidly with increasing order.

\section{Algebraic analysis}

\subsection{The interaction model}

In the ultraviolet, the kernel of the quark DSE, Eq.~(\ref{gap_eqn}), takes
the form
\begin{equation}
Z_{1}\gamma_{\mu}g^{2}D_{\mu\nu}(k)\Gamma_{\nu}(q,p)\rightarrow4\pi
\alpha(k^{2})\gamma_{\mu}D_{\mu\nu}^{\mathrm{free}}(k)\gamma_{\nu
},\label{LR_truncation}%
\end{equation}
where $k=p-q$, and $\alpha(k^{2})$ is the renormalized strong running
coupling, which has absorbed the renormalization constants of the quark and
gluon propagators and the vertex. The ladder-rainbow truncations that have been
phenomenologically successful in recent years for light quark hadrons adopt
the form of Eq.~(\ref{LR_truncation}) for all $k^{2}$ by replacing
$\alpha(k^{2})$ by $\alpha_{\mathrm{eff}}(k^{2})$, which contains the correct
1-loop QCD ultra-violet form and a parameterized infrared behavior fitted to
one or more chiral observables such as $\langle\bar{q}q\rangle_{\mu}^{0}$. In
this sense, such an $\alpha_{\mathrm{eff}}(k^{2})$ contains those infrared
effects of the dressed vertex $\Gamma_{\nu}(q,p)$ that can be mapped into a
single effective amplitude corresponding to $\gamma_{\nu}$ for chiral quarks.
Such a kernel does not have the explicit dependence upon quark mass that would
occur if the vertex dressing were to be generated by an explicit Feynman
diagram structure. In particular, one expects the vertex dressing to decrease
with increasing quark mass; the effective ladder-rainbow kernel appropriate to
heavy quark hadrons should have less infrared strength from dressing than is
the case for light quark hadrons.

We use an explicit (but approximate) diagrammatic description of the dressed
vertex $\Gamma_{\nu}(q,p)$, and to facilitate the analysis we make the
replacement $4\pi\alpha_{\text{eff}}(k^{2})/k^{2}\rightarrow(2\pi
)^{4}\mathcal{G}^{2}\delta^{4}(k)$. This is the Munczek-Nemirovsky
Ansatz~\cite{Munczek:1983dx} for the interaction kernel. The parameter
$\mathcal{G}^{2}$ is a measure of the integrated kernel strength, and we
expect this to be less than what would be necessary in ladder-rainbow format
because of the infrared structure now to be provided explicitly by the model
vertex $\Gamma_{\mu}(q,p)$. The equations of the previous Sections convert to
model form by the replacement
\begin{equation}
g^{2}D_{\mu\nu}(k)\rightarrow\left(  \delta_{\mu\nu}-\frac{k_{\mu}k_{\nu}%
}{k^{2}}\right)  (2\pi)^{4}\mathcal{G}^{2}\delta^{4}(k),
\label{EQ_GLUON_PROPAGATOR}%
\end{equation}
where we choose Landau gauge. It is the combination of
Eq.~(\ref{EQ_GLUON_PROPAGATOR}) and the model vertex that is the DSE kernel;
comparisons of Eq.~(\ref{EQ_GLUON_PROPAGATOR}) with information about the
dressed gluon 2-point function are incomplete. The resulting DSEs for the
quark propagator and gluon-quark vertex are ultra-violet finite; thus the
renormalization constants are unity: $Z_{1}=Z_{2}=1$, and there is no
distinction between bare and renormalized quark current mass.   We set 
\mbox{$m_{bm}=m(\mu)=m$}. 

\subsection{The algebraic vertex and quark propagator}

With this kernel, the vertex integral equation
Eq.~(\ref{EQ_IMPROVED_VERTEX_DSE_RAW}) determines solutions for $k=0$
and we define$~\Gamma_{\mu}(p,p):=\Gamma_{\mu}(p)$. The resulting
algebraic form for Eq.~(\ref{EQ_IMPROVED_VERTEX_DSE_RAW}) is%
\begin{equation}
\Gamma_{\mu}(p)=\gamma_{\mu}-\mathcal{CG}^{2}\Gamma_{\sigma}(p)S(p)\Gamma
_{\mu}(p)S(p)\Gamma_{\sigma}(p)~.\label{EQ_IMPROVED_VERTEX_DSE}%
\end{equation}
In obtaining this form, we have used $3{C_{\mathrm{F}}}/4=1$, where the extra
factor of $3/4$ arises from the transverse projector. The general form of the
vertex is:
\begin{multline}
\Gamma_{\mu}(p)=\alpha_{1}(p^{2})\gamma_{\mu}+\alpha_{2}(p^{2})\gamma\cdot
p~p_{\mu}-\alpha_{3}(p^{2})ip_{\mu}\label{EQ_VERTEX_FC_GEN}\\
+\alpha_{4}(p^{2})i\gamma_{\mu}~\gamma\cdot p
\end{multline}
where $\alpha_{i}(p^{2})$ are invariant amplitudes. From
Eq.~(\ref{EQ_IMPROVED_VERTEX_DSE}) we find $\alpha_{4}=0$, as was the case for
the related models in Refs. \cite{Bender:2002as} and \cite{Bhagwat:2004hn}.%
\begin{figure}
[ptb]
\begin{center}
\includegraphics[
height=3.2897in,
width=3.4126in
]%
{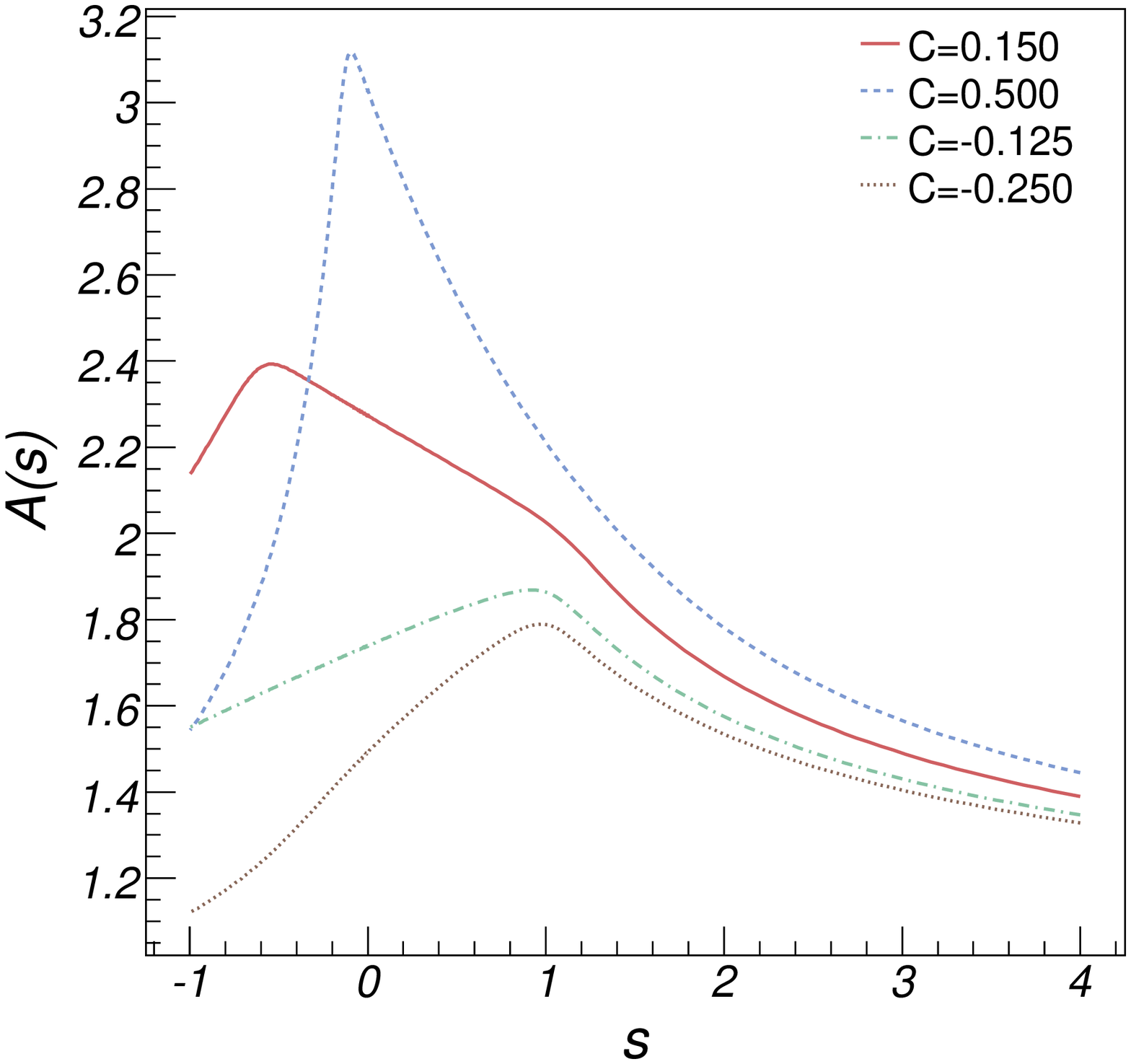}%
\caption{(Color Online) Quark propagator amplitude $A(s)$ versus Euclidean
$s=p^{2}$. We use the interaction mass scale $\mathcal{G}=1~\operatorname{GeV}%
$ and the current mass is $m=0.0183~\mathcal{G}=18.3\,$~$\operatorname{MeV}$.
$\mathcal{C}$ dependence calculated with converged summation of vertex
dressing, for $\mathcal{C}=0.15$ (solid curve), $\mathcal{C}=0.5$ (dashed
curve), $\mathcal{C}=-0.125$ (dot-dashed curve) and $\mathcal{C}=-0.25$(dotted
curve). }%
\label{PLOT_A_DIFF-CS}%
\end{center}
\end{figure}
\begin{figure}
[ptbptb]
\begin{center}
\includegraphics[
height=3.2897in,
width=3.4126in
]%
{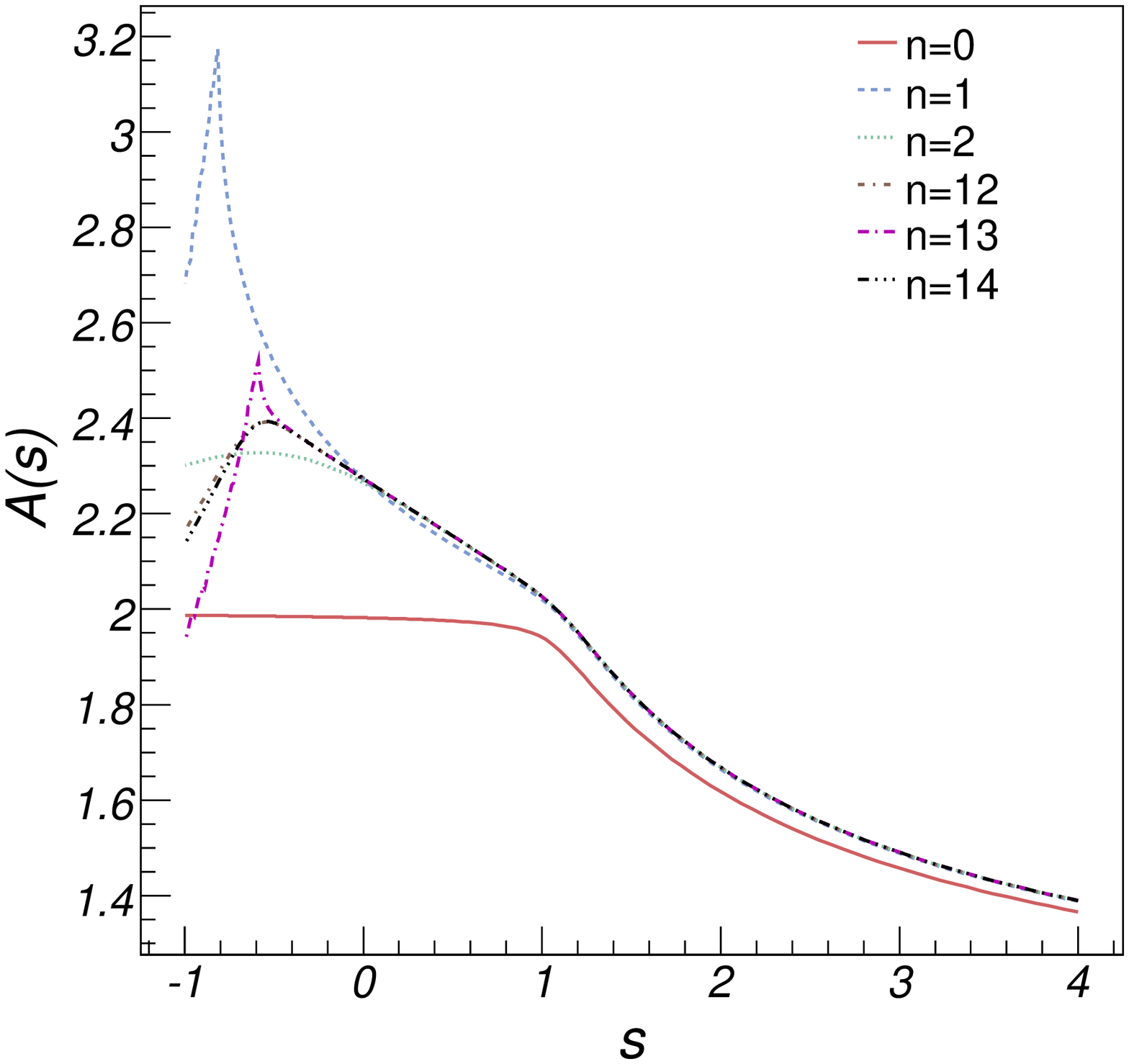}%
\caption{(Color Online) Quark propagator amplitude $A(s)$ versus Euclidean
$s=p^{2}$. We use the interaction mass scale $\mathcal{G}=1~\operatorname{GeV}%
$ and the current mass is $m=0.0183~\mathcal{G}=18.3\,$~$\operatorname{MeV}$.
We show the influence of vertex dressing to order $n$ as described in the
text. For $\mathcal{C}=0.15$, $n=0$ yields the solid curve and the result is
the ladder-rainbow truncation. The other curves are $n=1$ (long dashed curve,
1-loop vertex), $n=2$ (dotted curve, 2-loop vertex), $n=12$ (dot - short
dashed curve), $n=13$ (dot - long dashed curve) and $n=14$ (dot dot dot -
dashed curve) order of dressing of the quark gluon vertex.}%
\label{PLOT_A_DIFF-n}%
\end{center}
\end{figure}

The vertex is a sum over contributions with exactly $n$ internal effective
gluon kernels according to
\begin{equation}
\Gamma_{\mu}(p)=\sum_{n=0}^{\infty}\Gamma_{\mu}^{n}%
(p),\label{EQ_VERTEX_DECOMP}%
\end{equation}
with the general contribution given by the recursive relation
\begin{equation}
\Gamma_{\mu}^{n}(p)=-\mathcal{CG}^{2}\sum_{\substack{j,k,l\\n=j+k+l+1}%
}\Gamma_{\nu}^{j}(p)S(p)\Gamma_{\mu}^{k}(p)S(p)\Gamma_{\nu}^{l}%
(p),\label{EQ_VERTEX_RECUR}%
\end{equation}
where $\Gamma_{\mu}^{0}(p)=\gamma_{\mu}$. Substitution of the form
$S(p)^{-1}=$ $i\gamma\cdot pA(p^{2})+B(p^{2})$ into Eq.(\ref{EQ_VERTEX_RECUR})
gives $\Gamma_{\mu}^{n}(p^{2})$ in terms of the functions $A(p^{2})$ and
$B(p^{2})$. These latter functions must be solved simultaneously with the
vertex at the given order. The algebraic form of the gap equation for the
propagator is
\begin{equation}
S^{-1}(p)=i\gamma\cdot p+m +\mathcal{G}^{2}\gamma_{\mu}S(p)\Gamma_{\mu
}(p),\label{EQ_GAP_EQUATION_SIMPL}%
\end{equation}
where again the transverse projector and the color factor combine to yield
$3{C_{\mathrm{F}}}/4=1$. After projection onto the two Dirac amplitudes we
have
\begin{align}
A(p^{2}) &  =1-\mathcal{G}^{2}\frac{i}{4}tr\left[  \frac{\gamma\cdot p}{p^{2}%
}\gamma_{\mu}S(p)\,\Gamma_{\mu}(p)\right]  ,\label{EQ_PROPAGATOR_A}\\
B(p^{2}) &  =m+\mathcal{G}^{2}\frac{1}{4}tr\left[  \gamma_{\mu
}\,S(p)\,\Gamma_{\mu}(p)\right]  .\label{EQ_PROPAGATOR_B}%
\end{align}
Equations (\ref{EQ_VERTEX_RECUR}), (\ref{EQ_PROPAGATOR_A}) and
(\ref{EQ_PROPAGATOR_B}) are solved simultaneously at a specified order $n$ of
vertex dressing.%
\begin{figure}
[ptb]
\begin{center}
\includegraphics[
height=3.2897in,
width=3.4126in
]%
{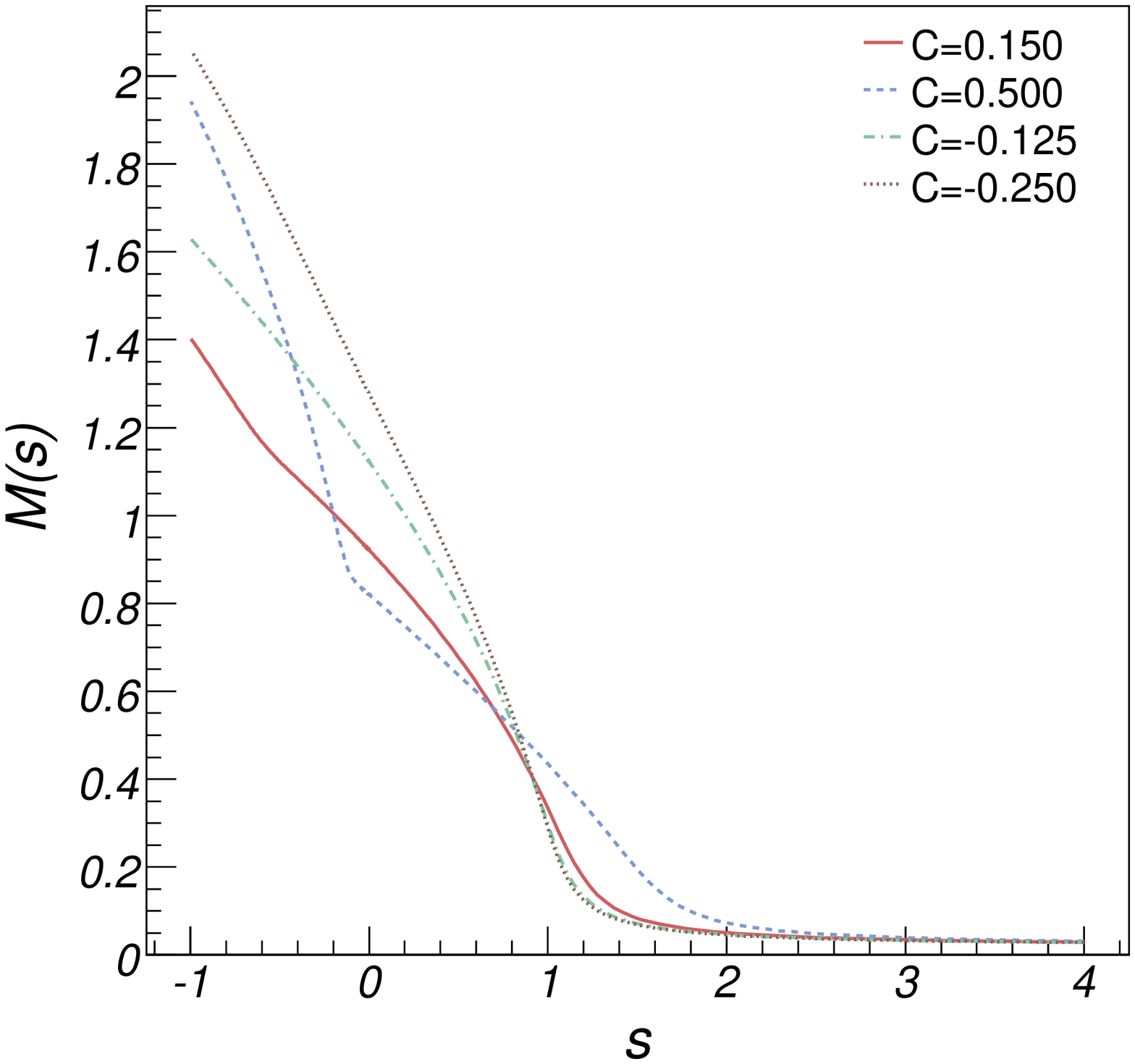}%
\caption{(Color Online) Quark mass function $M(s)$ versus Euclidean $s=p^{2}$.
We use the interaction mass scale $\mathcal{G}=1~\operatorname{GeV}$ and the
current mass is $m=0.0183~\mathcal{G}=18.3\,$~$\operatorname{MeV}$.
$\mathcal{C}$ dependence calculated with converged summation of vertex
dressing, for $\mathcal{C}=0.15$ (solid curve), $\mathcal{C}=0.5$ (dashed
curve), $\mathcal{C}=-0.125$ (dot-dashed curve) and $\mathcal{C}=-0.25$(dotted
curve). }%
\label{PLOT_M_DIFF-CS}%
\end{center}
\end{figure}
\begin{figure}
[ptbptb]
\begin{center}
\includegraphics[
height=3.2897in,
width=3.4126in
]%
{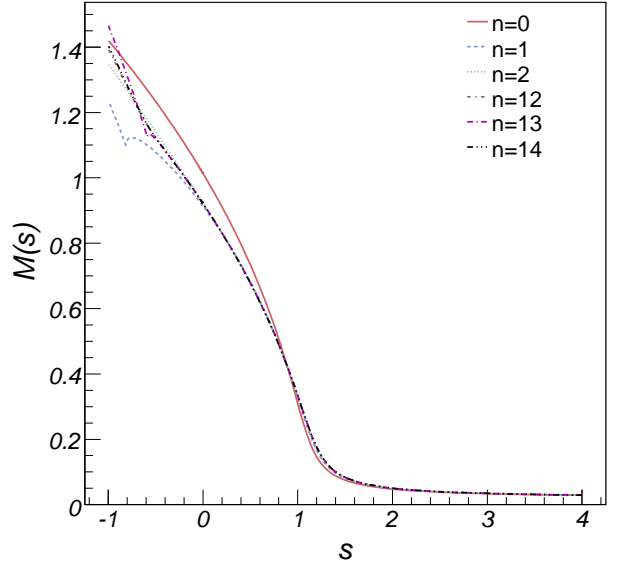}%
\caption{(Color Online) Quark mass function $M(s)$ versus Euclidean $s=p^{2}$.
We use the interaction mass scale $\mathcal{G}=1~\operatorname{GeV}$ and the
current mass is $m=0.0183~\mathcal{G}=18.3\,$~$\operatorname{MeV}$. We show
the influence of vertex dressing to order $n$ as described in the text. For
$\mathcal{C}=0.15$, $n=0$ yields the solid curve and the result is the
ladder-rainbow truncation. The other curves are $n=1$ (long dashed curve,
1-loop vertex), $n=2$ (dotted curve, 2-loop vertex), $n=12$ (dot - short
dashed curve), $n=13$ (dot - long dashed curve) and $n=14$ (dot dot dot -
dashed curve) order of dressing of quark gluon vertex.}%
\label{PLOT_M_DIFF-N}%
\end{center}
\end{figure}

In the case where one is limited to a strict ladder summation for the vertex 
with bare internal vertices, closed form expressions for the vertex amplitudes
$\alpha_{i}$ in terms of $A$ and $B$ are
obtainable~\cite{Bender:2002as,Bhagwat:2004hn}. With the enlarged class of
dressing considered here, corresponding closed form expressions have not been
obtained. However numerical evaluation is sufficient for the vertex and
propagator amplitudes; a numerical treatment of the BSE kernel must be made in
any case. Numerical solution of the simultaneous algebraic equations for the
vertex and propagator is carried out here using the algebraic and numerical
tools of \textit{Mathematica} \cite{Wolfram} with the assistance of the
\textit{FeynCalc} package used for computer-algebraic evaluation of the Dirac
algebra \cite{Mertig:1990an}.

The model parameter $\mathcal{C}$ for the vertex is determined by a fit to
selected global features of quenched lattice-QCD data for the quark
propagator~\cite{Bowman:2002kn} and the quark-gluon
vertex~\cite{Skullerud:2003qu}. This data is available for both quantities at
current quark mass $m=\bar{m}=60$~MeV. This is the same data as used to fit
the same parameter $\mathcal{C}$ in Ref.~\cite{Bhagwat:2004hn}; a different
result will therefore reflect the wider class of vertex dressing herein. To
facilitate comparison we also eliminate the role of the interaction strength
mass scale parameter $\mathcal{G}$ in this step by dealing with dimensionless
quantities; $\mathcal{G}$ will later be fixed by requiring that $m_{\rho}$ be reproduced.

The lattice-QCD data for the quark propagator indicates that $Z_{qu}(0)\equiv$
$1/A_{qu}(0)\approx0.7$ and $M_{qu}(0)\equiv B(0)/A(0)$ $\approx0.42$~GeV.
Following Ref.~\cite{Bhagwat:2004hn} the lattice data for both the propagator
and the vertex in the infrared is characterized by the set of four
dimensionless quantities evaluated at $p^{2}=0$:
\begin{equation}
A(0,m_{60})=1.4 \label{EQ_LATTICE_A}%
\end{equation}%
\begin{equation}
\alpha_{1}(0,m_{60})=2.1 \label{EQ_LATTICE_ALPHA1}%
\end{equation}%
\begin{equation}
-M(0,m_{60})^{2}\alpha_{2}(0,m_{60})=7.1 \label{EQ_LATTICE_ALPHA2}%
\end{equation}%
\begin{equation}
-M(0,m_{60})\alpha_{3}(0,m_{60})=1.0, \label{EQ_LATTICE_ALPHA3}%
\end{equation}
where $m_{60}=\bar{m}/M_{qu}(0)$. The best fit to these quantities gives
$\mathcal{C}=0.34$ with an average relative error of $\bar{r}=24$~\% and
standard deviation $\sigma_{r}=70$~\%. The quality of fit is about the same as
in Ref.~\cite{Bhagwat:2004hn}, and changes $\Delta\mathcal{C}\approx\pm0.2$
are not significant in this regard. For example, $\mathcal{C}=0.15$ leads to
$\bar{r}=39$~\% and $\sigma_{r}=72$~\%. We will use $\mathcal{C}=0.15$ because
the resulting vertex at timelike $p^{2}$ is more convergent with respect to
increasing order of dressing. The value of $\mathcal{C}$ being significantly
greater than the strict ladder sum limit $\mathcal{C}=-1/8$, we see that the
attraction provided by the 3-gluon coupling is important for the vertex.
However the amount of attraction that must be provided in this
phenomenological way in the present work is less than what was required in
Ref.~\cite{Bhagwat:2004hn} to fit the same lattice quantities. In that work,
$\mathcal{C}=0.51$ was found necessary. We attribute this difference to the
fact that a wider class of self-consistent dressing diagrams is included in
the present approach; attraction is provided by every vertex that is internal 
in the sense of Fig.~{\ref{PLOT_IMPROVED_VERTEX}}.

In Figs.~{\ref{PLOT_A_DIFF-CS} and \ref{PLOT_A_DIFF-n}} we present the results
for our calculations of $A(p^{2})$ for different values of $\mathcal{C}$ and
different orders of quark-gluon vertex dressing. We set $\mathcal{G}%
=1~\mathrm{GeV}$, so all dimensioned quantities are measured in units of
$\mathcal{G}$. The current mass is $m_{q}=0.0183~\mathcal{G}$. One can see
from Fig.~{\ref{PLOT_A_DIFF-CS}} that $\mathcal{C}$ has a major impact on the
behavior of $A(s)$, especially in the timelike region.
Fig.~{\ref{PLOT_A_DIFF-n}} shows that with $n=14$ as the order of dressing of
the quark-gluon vertex, we achieved convergence of the quark propagator
function $A(p^{2})$ for $p^{2}>-\mathcal{G}^{2}$. The same is true for the
function $B(p^{2})$. The relative measure of the convergence of the quark
propagator functions with $n$ is the convergence of the meson masses
calculated using the solutions for the propagators. It will be shown later on
that our calculations of $m_{\pi}$ and $m_{\rho}$ have converged to better
than $1\%$ for $n=14$. For heavier current quarks the convergence region for
the solutions of $A(p^{2})$ and $B(p^{2})$ extends deeper into the time-like
region of $p^{2}$, which allows for convergent calculations of heavier meson masses.

In Figs.~{\ref{PLOT_M_DIFF-CS}} and \ref{PLOT_M_DIFF-N} we present the results
for our calculations of $M(p^{2})=B(p^{2})/A(p^{2})$ for different values of
$\mathcal{C}$ and different orders of quark-gluon vertex dressing. Again these
calculations have $\mathcal{G}=1~\mathrm{GeV}$, so all dimensioned quantities
are measured in units of $\mathcal{G}$. The vertex parameter $\mathcal{C}$ has
a modest impact on the behavior of $M(s)$.

\begin{figure}
[ptb]
\begin{center}
\includegraphics[
height=3.2897in,
width=3.4126in
]%
{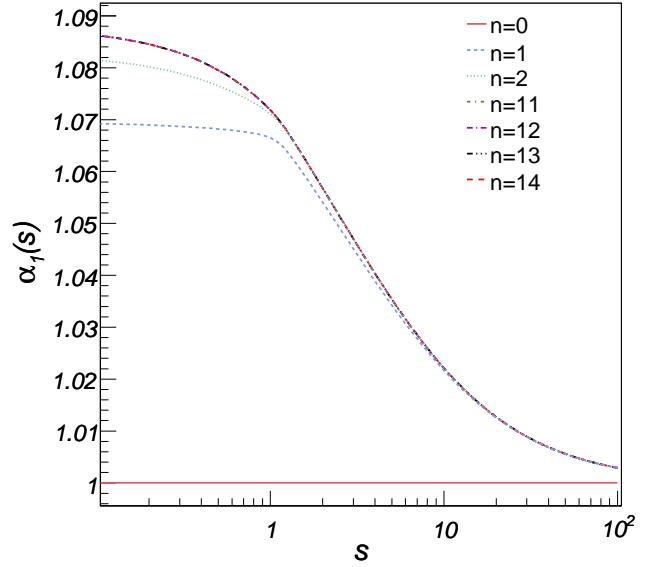}%
\caption{(Color Online) Gluon-quark vertex amplitude $\alpha_{1}(s)$ versus
Euclidean $s=p^{2}$, for $\mathcal{C}=0.15$. We use the interaction mass scale
$\mathcal{G}=1~\operatorname{GeV}$ and the current mass is
$m=0.0183~\mathcal{G}=18.3\,$~$\operatorname{MeV}$. $n=0$ (solid curve)
results from the bare vertex and is the ladder-rainbow truncation. The other
curves are $n=1$ (short dashed curve, 1-loop vertex dressing), $n=2$ (dotted
curve), $n=11$ (dot - short dashed curve), $n=12$ (dot - long dashed curve),
$n=13$ (dot dot dot - dashed curve) and $n=14$ (long dashed curve). }%
\label{PLOT_ALPHA_1}%
\end{center}
\end{figure}

\begin{figure}
[ptbptb]
\begin{center}
\includegraphics[
height=3.2897in,
width=3.4126in
]%
{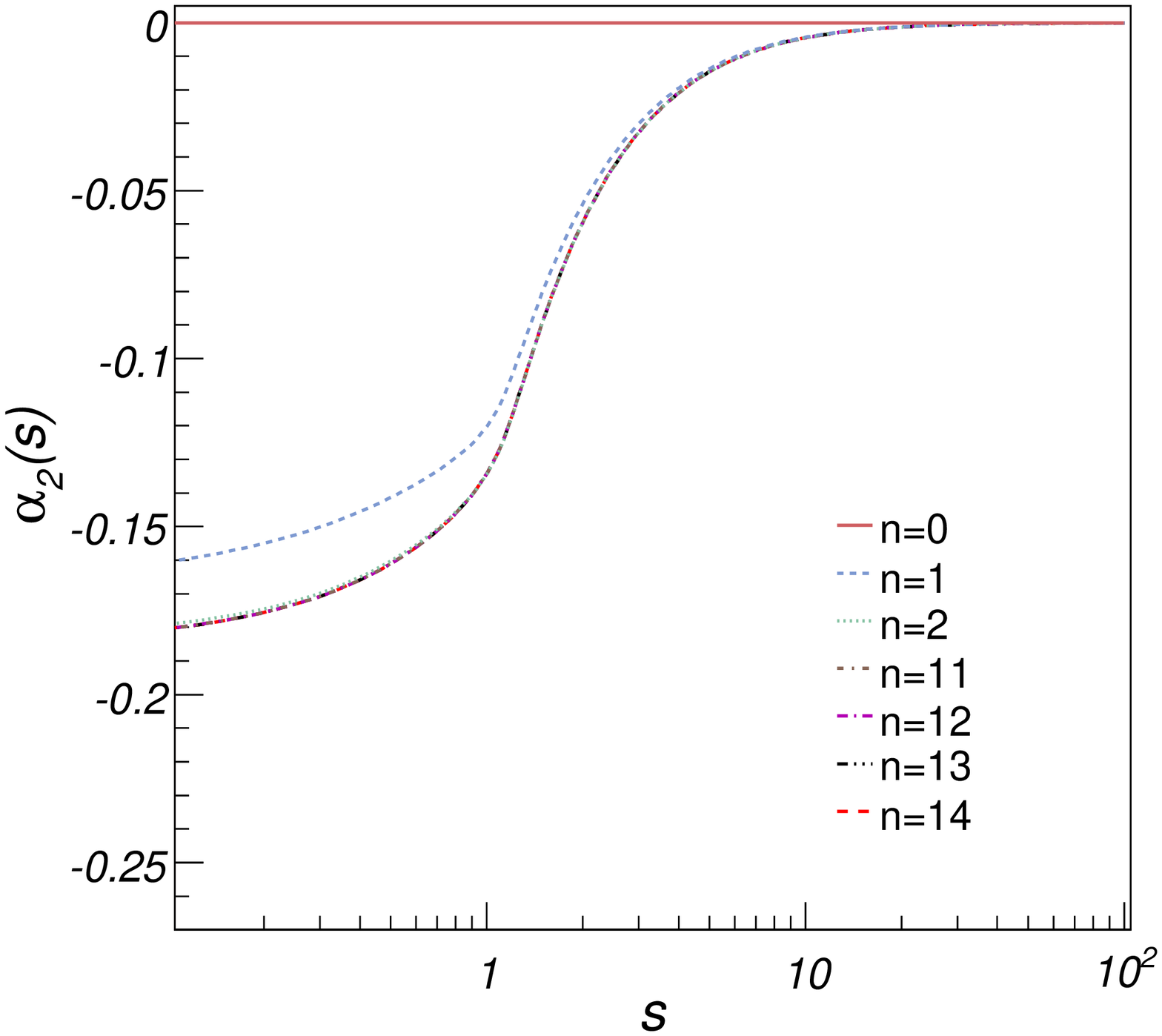}%
\caption{(Color Online) Gluon-quark vertex amplitude $\alpha_{2}(s)$ versus
Euclidean $s=p^{2}$, for $\mathcal{C}=0.15$. We use the interaction mass scale
$\mathcal{G}=1~\operatorname{GeV}$ and the current mass is
$m=0.0183~\mathcal{G}=18.3\,$~$\operatorname{MeV}$. $n=0$ (solid curve)
results from the bare vertex and is the ladder-rainbow truncation. The other
curves are $n=1$ (short dashed curve, 1-loop vertex dressing), $n=2$ (dotted
curve), $n=11$ (dot - short dashed curve), $n=12$ (dot - long dashed curve),
$n=13$ (dot dot dot - dashed curve) and $n=14$ (long dashed curve). }%
\label{PLOT_ALPHA_2}%
\end{center}
\end{figure}

\begin{figure}
[ptbptbptb]
\begin{center}
\includegraphics[
height=3.2897in,
width=3.4126in
]%
{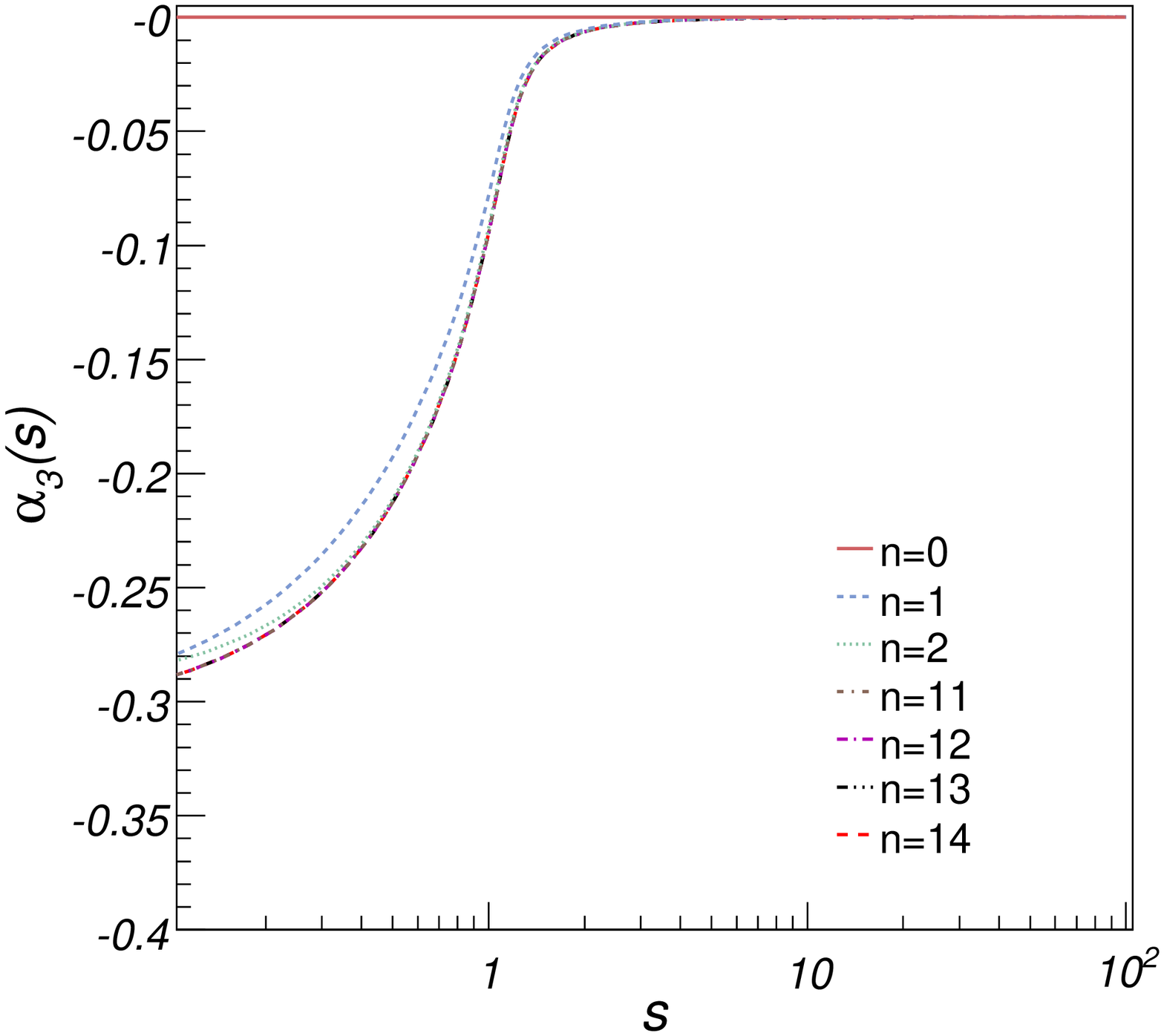}%
\caption{(Color Online) Gluon-quark vertex amplitude $\alpha_{3}(s)$ versus
Euclidean $s=p^{2}$, for $\mathcal{C}=0.15$. We use the interaction mass scale
$\mathcal{G}=1~\operatorname{GeV}$ and the current mass is
$m=0.0183~\mathcal{G}=18.3\,$~$\operatorname{MeV}$. $n=0$ (solid curve)
results from the bare vertex and is the ladder-rainbow truncation. The other
curves are $n=1$ (short dashed curve, 1-loop vertex dressing), $n=2$ (dotted
curve), $n=11$ (dot - short dashed curve), $n=12$ (dot - long dashed curve),
$n=13$ (dot dot dot - dashed curve) and $n=14$ (long dashed curve). }%
\label{PLOT_ALPHA_3}%
\end{center}
\end{figure}

Figs. {\ref{PLOT_ALPHA_1}, \ref{PLOT_ALPHA_2}} and {\ref{PLOT_ALPHA_3}}
display the results for the vertex amplitudes $\alpha_{i}(s)$ corresponding to
different orders of vertex dressing. Successive orders after 1-loop ($n=1$)
serve to enhance the infrared strength for $s<1$. The convergence with $n$ is
monotonic, in contrast to the convergence of the BSE kernel that is generated from
this vertex, as discussed later.

The quark condensate in the present model is given by
\begin{equation}
\langle\bar{q}q\rangle^{0}=-\frac{3}{4\pi^{2}}\int_{0}^{s_{0}}dss\frac
{B_{0}(s)}{s\,A_{0}^{2}(s)+B_{0}^{2}(s)},
\end{equation}
in terms of the chiral limit quark propagator amplitudes. There is no
renormalization necessary because there is a spacelike $s_{0}$ for which
$B_{0}(s>s_{0})=0$. Because of the underrepresentation of the ultra-violet
strength of the interaction in this model, the condensate is
characteristically too low. In particular we find
\begin{equation}
-\langle\bar{q}q\rangle_{\mathcal{C}=0.15}^{0}=(0.2146~\mathcal{G}%
)^{3}=(0.1266~\mathrm{GeV})^{3}\label{cond_full}%
\end{equation}
with $\mathcal{G}=0.59$~GeV. The rainbow-ladder result ($\mathcal{C}=0$) is
$-\langle\bar{q}q\rangle_{\mathrm{LR}}^{0}=\mathcal{G}^{3}/(10\pi
^{2})=(0.1277~\mathrm{GeV})^{3}$. Thus one can see that the vertex dressing
decreases the condensate slightly. In more detail, we have
\begin{equation}
\frac{\langle\bar{q}q\rangle_{\mathrm{LR}}^{0}}{\langle\bar{q}q\rangle
_{\mathcal{C}=0.15}^{0}}=1.03,
\end{equation}
which indicates that the ladder-rainbow truncation overestimates the
condensate by 3\% compared to the more completely dressed vertex considered
here. The previous study~\cite{Bhagwat:2004hn} with a more restricted class of
vertex dressing diagrams found that the ladder-rainbow truncation was 18\% too
low.%

\subsection{The algebraic Bethe-Salpeter kernel for mesons}

Substitution of the model interaction kernel Eq.~(\ref{EQ_GLUON_PROPAGATOR})
into the meson BSE, Eq.~(\ref{EQ_BSE_MODEL}), produces the algebraic form
\begin{multline}
\Gamma_{M}(k;P)=-\mathcal{G}^{2}\gamma_{\mu}\left\{  \chi_{M}(k;P)\Gamma_{\mu
}(k_{-})\right.  \label{EQ_BSE_MODEL_MN}\\
\left.  +S(k_{+})\Lambda_{M\mu}(k;P)\right\}  .
\end{multline}
The previous general recurrence relation Eq.~(\ref{EQ_LAMBDA_RECUR}) for the
general term of $\Lambda_{M\nu}=\sum_{n=0}^{\infty}\Lambda_{M\nu}^{n}$ now has
the algebraic form
\begin{gather}
\Lambda_{M\nu}^{n}(k;P)=-{\mathcal{C}}\mathcal{G}^{2}\sum
_{\substack{j,k,h\\n=j+k+h+1}}\nonumber\\
\left[  \Gamma_{\rho}^{j}(k_{+})\chi_{M}(k;P)\Gamma_{\nu}^{k}(k_{-}%
)\,S(k_{-})\Gamma_{\rho}^{h}(k_{-})\right.  \nonumber\\
\left.  +\Gamma_{\rho}^{j}(k_{+})S(k_{+})\Gamma_{\nu}^{k}(k)\chi
_{M}(k;P)\Gamma_{\rho}^{h}(k_{-})\right.  \nonumber\\
\left.  +\Lambda_{M\rho}^{j}(k;P)S(k_{-})l^{a}\Gamma_{\nu}^{k}(k_{-}%
)S(k_{-})\Gamma_{\sigma}^{h}(k_{-})\right.  \nonumber\\
\left.  +\Gamma_{\rho}^{j}(k_{+})S(q_{+})\Lambda_{M\nu}^{k}(k;P)S(k_{-}%
)\Gamma_{\rho}^{h}(k_{-})\right.  \nonumber\\
\left.  +\Gamma_{\rho}^{j}(k_{+})S(k_{+})\Gamma_{\nu}^{k}(k_{+})S(k_{+}%
)\Lambda_{M\sigma}^{h}(k;P)\right]  .\label{EQ_LAMBDA_MN}%
\end{gather}

If we work at a given order, $n$, of vertex dressing, then the quark
propagator, the dressed vertex, and the BSE kernel can be constructed
recursively. By construction, chiral symmetry is preserved and the chiral
pseudoscalar states are massless, independent of model parameters. Because of
the algebraic structure, in which \ the BS amplitude $\Gamma_{M}(k;P)$ appears
on both sides of Eq.~(\ref{EQ_BSE_MODEL_MN}) with the same $q\bar{q}$ relative
momentum $k$, a physical solution where $P^{2}=-M_{M}^{2}$ is independent of
$k$ is defined only at $k=0$. That is, the quark and antiquark have momenta
$\eta P$ and $(1-\eta)P$. (Here we consider only equal mass quarks and thus
have chosen $\eta=1/2$.) The Munczek-Nemirovsky model interaction does not
allow momentum transfer to quarks. This is a restriction present in all hadron
studies made within this model. We define $\Gamma_{M}(P)=\Gamma_{M}(k=0;P)$,
after which the form in which we solve the BSE is
\begin{multline}
\Gamma_{M}(P)=-\mathcal{G}^{2}\,\gamma_{\mu}S(\frac{P}{2})\left\{  \Gamma
_{M}(P)\,S(-\frac{P}{2})\,\Gamma_{\mu}(-\frac{P}{2})\right.
\label{EQ_BSE_MODEL_0}\\
\left.  +\Lambda_{M\mu}(0;P)\right\}  .
\end{multline}

\section{Meson Masses and Results}

The general form of a meson BS amplitude can be written as
\begin{equation}
\Gamma_{M}(k;P)=\sum_{i}\mathcal{K}^{i}(k;P)\,f_{M}^{i}(k^{2},k\cdot
P;P^{2}),\label{EQ_BSA_MESON}%
\end{equation}
where the $\mathcal{K}^{i}(k;P)$ are a complete set of independent covariants
constructed from Dirac matrices and momenta that transform in a manner
specified by the quantum numbers of the meson under consideration. The
$f_{M}^{i}(k^{2},k\cdot P;P^{2})$ are the corresponding invariant amplitudes.
(We do not show explicitly the color singlet unit matrix.) The model BSE under
consideration here, Eq.~(\ref{EQ_BSE_MODEL_0}), has relative momentum $k=0$,
and the set of covariants is reduced considerably. We have
\begin{equation}
\Gamma_{M}(P)=\sum_{i=1}^{N}\,\mathcal{K}^{i}(P)f_{M}^{i}(P^{2}%
)\label{EQ_BSA_MODEL}%
\end{equation}
and it is convenient to develop a set of projection operators $\mathcal{P}%
_{j}$ that allow us to isolate each amplitude according to
\begin{equation}
f_{M}^{j}=Tr_{D}\left[  \mathcal{P}_{j}\Gamma_{M}\right]
.\label{EQ_H_PROJECTOR}%
\end{equation}
Then projection of the BSE, Eq.~(\ref{EQ_BSE_MODEL_0}), yields the eigenvalue
equation
\begin{equation}
f(P^{2})=\mathcal{H}(P^{2})\,f(P^{2}),\label{EQ_F_SCAL_FUNC}%
\end{equation}
where $f=(f_{M}^{1},f_{M}^{2}\cdots)$ is a vector of invariant amplitudes and
the matrix $\mathcal{H}(P^{2})$ is an $N\times N$ representation of the kernel.

The mass, $M_{M}$, of the lowest bound state is obtained from the highest
negative value of $P^{2}$ for which
\begin{equation}
\det\left[  \mathcal{H}(P^{2})-I\right]  _{P^{2}+M_{M}^{2}=0}%
=0.\label{EQ_MESON_CHAR_POLYNOMIAL}%
\end{equation}
This method, namely the solution of the characteristic polynomial for
Eq.~(\ref{EQ_F_SCAL_FUNC}), has also been followed in earlier work of this
type~\cite{Bender:2002as,Bhagwat:2004hn}.

\subsection{The Pion}

The general form of the $\pi$ Bethe-Salpeter amplitude requires four
covariants and is
\begin{equation}
\Gamma_{\pi}(k;P)=\gamma_{5}\,[i\,f_{\pi}^{1}+\gamma\cdot P\,f_{\pi}%
^{2}+\gamma\cdot k\,k\cdot P\,f_{\pi}^{3}+\sigma_{\mu\nu}k_{\mu}P_{\nu
}\,f_{\pi}^{4}], \label{EQ_BSA_PION_GEN}%
\end{equation}
in terms of amplitudes $f_{\pi}^{i}(k^{2},k\cdot P;P^{2})$. We do not show
flavor dependence since we treat we treat u-quarks and d-quarks the same in
all other respects. In the present case only two covariants survive and we
have
\begin{equation}
\Gamma_{\pi}(P)=\gamma_{5}\,[if_{\pi}^{1}(P^{2})+\gamma\cdot Pf_{\pi}%
^{2}(P^{2})]. \label{EQ_BSA_PION_MODEL}%
\end{equation}
Convenient projection operators in this case are
\begin{equation}
\mathcal{P}_{1}=-\frac{i}{4}\gamma_{5},~\ \mathcal{P}_{2}=\frac{1}{4P^{2}%
}\gamma\cdot P\gamma_{5}. \label{EQ_BSA_PION_PROJECTOR}%
\end{equation}

\subsection{The Rho}

The general form of the $\rho$ Bethe-Salpeter amplitude requires eight
transverse covariants and corresponding amplitudes. Specific choices that have
been found convenient in earlier work are given in
Refs.~\cite{Maris:1999nt,Jarecke:2002xd}. In the present case, the most
general form is simply
\begin{equation}
\Gamma_{\rho\,\mu}(P)=\left(  \delta_{\mu\nu}-\frac{P_{\mu}P_{\nu}}{P^{2}%
}\right)  \,\gamma_{\nu}f_{\rho}^{1}(P^{2})+\sigma_{\mu\nu}P_{\nu}f_{\rho}%
^{2}(P^{2}).\label{EQ_BSA_RHO_MODEL}%
\end{equation}
Again a unit color matrix is understood and we treat u-quarks and d-quarks as
the same. Convenient projection operators that isolate the amplitudes are
\begin{equation}
\mathcal{P}_{1}=\frac{1}{12}\gamma_{\mu},~\ \mathcal{P}_{2}=\frac{1}{12P^{2}%
}\sigma_{\mu\nu}P_{\nu}.\label{EQ_BSA_RHO_PROJECTOR}%
\end{equation}

\subsection{Vertex Dressing for Light Quarks}

There are a total of three parameters: $\mathcal{C}=0.15$, which has already
been set by the quenched lattice data for the quark propagator and the
gluon-quark vertex, while the experimental $m_{\pi}$ and $m_{\rho}$ are used
to set the other two: the interaction mass scale, $\mathcal{G}=0.59$~GeV, and
the current mass for the u/d quark, $m=0.0183\,\mathcal{G}=11\,$~MeV. The
fully dressed vertex model is used in these determinations. In practice, we
require convergence to 3 significant figures for the masses and this is
achieved with a vertex dressed to order $n=14$. Table~{\ref{table_rho_dress}}
shows how the vertex dressing influences $m_{\pi}$ and $m_{\rho}$.
\begin{table}[t]
\caption{Effect of quark-gluon vertex dressing to order $n$ upon the masses of
the $\pi$ and $\rho$ mesons (in GeV). The ladder-rainbow (LR) truncation
corresponds to $n=0$, one loop vertex dressing corresponds to $n=1$, etc, while
the full model result (converged to 3 significant figures) is labeled
$n=\infty$. Also displayed for $m_{\rho}$ is the mass error, $\Delta m_{\rho}$,
and the relative mass error, $\Delta m_{\rho}/m_{\rho}$, of the LR truncation of
the present model compared to a previous model~\cite{Bhagwat:2004hn} based on
a limited class of vertex dressing diagrams. The mass scale parameter is
$\mathcal{G}=0.59$~GeV, the current mass of the u/d-quark is
$m=0.0183\,\mathcal{G}=11\,$~MeV, and $\mathcal{C}=0.15$. \vspace*{1ex} }%
\label{table_rho_dress}
\begin{ruledtabular}
\begin{tabular*}
{\hsize} {l@{\extracolsep{0ptplus1fil}}
|c@{\extracolsep{0ptplus1fil}}c@{\extracolsep{0ptplus1fil}}
c@{\extracolsep{0ptplus1fil}}c@{\extracolsep{0ptplus1fil}}
|c@{\extracolsep{0ptplus1fil}}}
Vertex Dressing  & $m_\pi$ & $m_\rho$ & $\Delta m_\rho$ & $\frac{\Delta m_\rho}{m_\rho}$ &
$\frac{\Delta m_\rho}{m_\rho}$~\protect\cite{Bhagwat:2004hn} \\\hline
$n=0$ (LR)      & 0.140       & 0.850       &  +0.074  & +0.095 & +0.295      \\
$n=1$ (1-loop)   & 0.135      & 0.759       &  -0.017  & -0.022 &-----\\
$n=2$            & 0.135      & 0.781       & +0.005   & +0.006 & +0.096      \\
$n=3$            & 0.135      & 0.772       & -0.004   & -0.005 &  N/A       \\
$n=4$            & 0.135      & 0.778       & +0.002   & +0.003 &  N/A      \\
$n=\infty$ (full model) & 0.135 & 0.776     & 0.0      & 0.0    &   0.0     \\
\end{tabular*}
\end{ruledtabular}\end{table}

To confirm that our constructed BSE kernel preserves chiral symmetry, we
verified that to any order of vertex dressing, and with $m=0$, the chiral pion
is massless to the numerical accuracy considered. The physical $m_{\pi}$ is
not fixed perfectly by the symmetry but is almost so. The explicit symmetry
breaking by the current mass is sufficient to determine $m_{\pi}$ for all
orders of vertex dressing except for a few \% error in the ladder-rainbow
truncation ($n=0$). Since the same behavior was observed in earlier work of
this nature~\cite{Bender:2002as,Bhagwat:2004hn}, this result is quite model-independent.

The response of $m_{\rho}$ to increasing order of vertex dressing shows that
the ladder-rainbow truncation is missing 74~MeV of attraction compared to the
full model result. The error decreases with each added order of vertex
dressing. The relative error in the ladder-rainbow mass is 9.5\% in the
present vertex model, compared to 29.5\% in the vertex model of
Ref.~\cite{Bhagwat:2004hn}.  Here each diagram for the dressed vertex has each 
of its internal vertices dressed in a self-consistent way. There are evidently 
both attractive and repulsive contributions at the various orders that combine 
to add less net attraction to the ladder-rainbow truncation than
what was found in Ref.~\cite{Bhagwat:2004hn}. 

\begin{table}[t]
\caption{Error of the ladder-rainbow truncation for equal quark mass vector mesons
in the u/d-, s-, c-, and b-quark regions, according to calculated mass and
effective binding energies (in GeV). The ladder-rainbow (LR) truncation
corresponds to order $n=0$ in vertex dressing and the full model result
corresponds to vertex dressing to all orders, $n=\infty$, in this model. The
mass scale parameter is $\mathcal{G}=0.59$~GeV, and $\mathcal{C}=0.15$. }%
\label{table_binding}%
\begin{ruledtabular}
\begin{tabular}{l|cccc}
&  ladder-rainbow  &  full model  & \multicolumn{2}{c}{LR \% error}  \\
&    $n=0$         &  $n=\infty$  &   this model  &  \protect\cite{Bhagwat:2004hn}\\ \hline
$m_{u,d}=0.011$ & & & &  \\
$m_\rho$                   & 0.850 & 0.776  & 9.5\%  & 30\%  \\
$\mathcal{BE}_\rho$        & 0.346 & 0.311  & 11\%   &        \\
$m_s=0.165$           & & & &  \\
$m_\phi$                    & 1.08  & 1.02   & 6.0\%  & 21\%   \\
$\mathcal{BE}_\phi$        & 0.350 & 0.320  & 9.0\%  &        \\
$m_c = 1.35$    & & & &  \\
$m_{J/\psi}$               & 3.11  & 3.09   & 0.3\%  & 3.5\%   \\
$\mathcal{BE}_{J/\psi}$   & 0.260 & 0.260  & 0\%    &        \\
$m_b=4.64$      & & & &  \\
$m_\Upsilon$               & 9.46  & 9.46   & 0\%    &  0\%   \\
$\mathcal{BE}_\Upsilon$   & 0.100 & 0.100  & 0\%    &        \\
%
\end{tabular}
\end{ruledtabular}
\end{table}

\subsection{Current Quark Mass Dependence}

One expects the influence of vertex dressing to decrease with increasing quark
mass because of the internal quark propagators in the vertex. Thus the LR
truncation should become more accurate for mesons involving heavier quarks. It
is useful to quantify this for the following reason. Phenomenological LR
kernels~\cite{Maris:2000sk} are capable of incorporating many realistic
features of QCD modeling and have been developed to provide efficient
descriptions of light quark mesons, their elastic and transition form factors,
and decay constants. A parameterized LR kernel that reproduces the
experimental $m_{\pi}$ and $m_{\rho}$ has, by definition, absorbed the
effective dressing of the vertex. The present work suggests that this is an
amount of vertex attraction worth $9.5\%$ of the vector meson mass. However
this phenomenological representation of the dressing does not have an explicit
dependence upon quark mass that would occur if the vertex dressing were to be
generated by an explicit Feynman diagram structure. One would expect such a
phenomenological LR kernel to be progressively too attractive when applied to
meson with progressively heavier quarks.

The present model provides an opportunity to explore how much of the final
meson mass result is attributable to vertex dressing and how this varies with
quark mass. In Table~{\ref{table_binding}} we display results for the ground
state vector mesons in the u/d-, s-, c-, and b-quark regions for both
rainbow-ladder truncation and the full model. Again the quark current masses
are determined so that the full model reproduces experiment. We see that the
amount by which the LR masses are too large decreases steadily with increasing
quark mass, as expected. The LR truncation here is missing $6\%$ of attraction
for $m_{\phi}$ compared to $21\%$ in the restricted class of dressing diagram
considerably previously~\cite{Bhagwat:2004hn}. The LR truncation is quite 
accurate for the $c\bar{c}$ and $b\bar{b}$ vector states, as expected.

For the larger quark masses, the meson mass is dominated by the sum of the
quark masses. We also express the results in a form that has this large mass
scale removed. For each state in Table~{\ref{table_binding}}, we display an
effective binding energy defined as $\mathcal{BE}=2M_{q}(0)-m_{V}$, where
$M_{q}(0)$ is the quark mass function obtained from the DSE solution at
$p^{2}=0$, and $m_{V}$ is the meson mass. Thus $M_{q}(0)$ is being used as a
rough measure of the constituent quark mass. The use of a single $p^{2}$ point
may well be an overestimate of constituent masses. Furthermore, our fitted
current quark masses are on the upper edge of what is usually quoted at a
renormalization scale of $\mu=2$~GeV~\cite{PDG04}. Such an overestimate would
be amplified in the infrared region via a DSE solution for the quark
propagator. Nevertheless, a relative comparison should be meaningful.
Table~{\ref{table_binding}} shows the dependence of $\mathcal{BE}$ upon the
current quark mass for the fully dressed model and the ladder-rainbow
truncation. On this basis, the relative amount of overbinding of the LR
truncation is consistent with its relative lack of attraction with respect to
the mass results.

\begin{table}[tb]
\caption{The masses of the equal quark mass vector and pseudoscalar mesons in the
u/d-, s-, c-, and b-quark regions, and the current quark masses required to
reproduce the experimental vector meson masses. All are in units of GeV. The
values of $m_{\eta_{c}}$ and $m_{\eta_{b}}$ are predictions.
Experimentally~\cite{PDG04}, $m_{\eta_{c}}=2.9797\pm0.00015\,$ and
$m_{\eta_{b}}=9.30\pm0.03$. The fictitious pseudoscalar $0_{s\bar{s}}^{-}$ is
included for comparison with other studies~\cite{Bhagwat:2004hn}.
\vspace*{1ex}}%
\label{table_mesons}%
\begin{ruledtabular}
\begin{tabular*}
{\hsize} {r@{\extracolsep{0ptplus1fil}}
r@{\extracolsep{0ptplus1fil}}
r@{\extracolsep{0ptplus1fil}}r@{\extracolsep{0ptplus1fil}}}
$m_{u,d}=$ 0.011 & $m_s=$ 0.165     & $m_c =$ 1.35         & $m_b=$ 4.64 \\
$m_\rho = 0.776$ & $m_\phi = 1.02$   & $m_{J/\psi} = 3.09$ & $m_{\Upsilon(1S)}=9.46$\\
$\mathcal{BE}_\rho=0.311$ & $\mathcal{BE}_\phi=0.320$ & $\mathcal{BE}_{J/\psi}=0.260$ &
$\mathcal{BE}_\Upsilon=0.100$ \\ \hline
$m_\pi = 0.135$  & $m_{0^-_{s\bar s}} = 0.61$ & $m_{\eta_c}=2.97$ & $m_{\eta_b}=9.43$    \\
$\mathcal{BE}_\pi=0.953$ & $\mathcal{BE}_{0^-}=0.727$ & $\mathcal{BE}_{\eta_c}=0.380$ &
$\mathcal{BE}_{\eta_b}=0.130$ \\
\end{tabular*}
\end{ruledtabular}
\end{table}

In Table~{\ref{table_mesons}} we display the full model results for both the
vector and pseudoscalar $q\bar{q}$ states. The masses for $\eta_{c}$ and
$\eta_{b}$ are predictions. In the c- and b-quark regions, these results are
essentially the same as those of Ref.~\cite{Bhagwat:2004hn}, because the
differences in the employed model of vertex dressing become irrelevant when
any dressing contribution is suppressed by the large mass of propagators
internal to the vertex. The systematics of the mass dependence of hyperfine
splitting that spans the c- and b-quark regions, here and in earlier
work~\cite{Bhagwat:2004hn}, strongly suggests that the experimental
value~\cite{PDG04}, $m_{\eta_{b}}=9.30\pm0.03$, is too low.

\section{Summary}

We have taken advantage of an algebraic model to enlarge the class of diagrams
for the quark-gluon dressed vertex that can be incorporated into the
Bethe-Salpeter kernel, while allowing a practical application to the
calculation of meson masses. A given expansion of the vertex in diagrammatic
form, produces a diagrammatic expansion of the quark self-energy, which in
turn specifies a diagrammatic expansion of the BSE kernel if chiral symmetry
is to be respected. This procedure relieves the phenomenology of the task of
reproducing Goldstone's theorem whenever parameters are changed - it is always
obeyed in this approach and thus phenomenology can address itself to a more
constrained task. The constraints are considerable: a realistic ladder-rainbow
kernel fitted to $\langle\bar{q}q\rangle^{0}$ \cite{Maris:2000sk} produces
$m_{\rho}$, $m_{\phi}$ and $m_{K^{\star}}$ to better than $5\%$. Such a
phenomenological LR kernel for light mesons has absorbed vertex dressing but
without the explicit $m_{q}$ dependence associated with an explicit
diagrammatic representation of the dressed gluon-quark vertex. In order to
gain more information it is necessary to work with a model that can implement
a summation of vertex diagrams, turn that into a summation of diagrams for the
chiral symmetry preserving BSE kernel, and allow a practical solution of the
meson BSE.

To this end we use the Munczek-Nemirovsky Ansatz ~\cite{Munczek:1983dx} for
the interaction kernel. We use an improved model for the quark-gluon dressed
vertex wherein each diagram for the dressed vertex has each of its internal
vertices dressed in a self-consistent way. This moves considerably beyond the
ladder BSE structure ~\cite{Bhagwat:2004hn} for the vertex, in which vertices
internal to the dressed vertex of interest are bare. In common with
Ref.~\cite{Bhagwat:2004hn}, we also use an effective method, with one
parameter ($\mathcal{C}=0.15$ for this model), to accommodate the important
non-Abelian effect of the 3-gluon coupling for the vertex. Quenched
lattice-QCD data for the quark propagator and the quark-gluon vertex at zero
gluon momentum fixed the parameter $\mathcal{C}$, while $m_{\pi}$ and
$m_{\rho}$ fixed the other two parameters via the fully dressed vertex results.

The resulting model provides a laboratory within which the relevance of
ladder-rainbow truncation (bare vertex) can be explored over a range of quark
masses from u/d-quarks to b-quarks. The influence of the enlarged class of
vertex dressing diagrams included in this work is seen to indicate that LR
truncation is missing 9.5\% of attraction for $m_{\rho}$, whereas the previous
information from a smaller class of vertex dressing diagrams had LR missing
30\% of attraction. The extra dressing diagrams included here tend to provide
some cancellation, making the LR truncation somewhat more accurate. As heavier
$q\bar{q}$ mesons are considered, the amount of missing attraction in the LR
truncation decreases steadily, as does the influence of vertex dressing---it
is less than $1\%$ for the $J/\psi$ and $\Upsilon$.

The influence of the non-Abelian 3-gluon coupling is very significant. No
attempt has been made to consider 4-gluon coupling nor to consider non-planar
gluon line diagrams (e.g. crossed-box diagrams) for the vertex dressing. 
On the other hand, a limited class of non-planar gluon line diagrams for 
the meson BSE kernel, as generated from the planar diagrams of
the dressed vertex, are included.  While the complex task of
including both planar and non-planar two-point gluon line diagrams for the
vertex is currently underway, it is not yet known whether explicit 3- and 
4-gluon couplings can be accommodated, even through the device of an effective
color factor.

\begin{acknowledgments}
This work was supported in part by DOE grant DE-AC05-84ER40150, under which
SURA operates Jefferson Lab, the U.S. National Science Foundation under Grant
Nos. PHY-0500291 and PHY-0301190 and the Southeastern Universities Research
Association(SURA). HHM thanks Jerry P. Draayer for his support during the
course of the work, the Graduate School of Louisiana State University for a
fellowship partially supporting his research, George S. Pogosyan and Sergue I.
Vinitsky for their support at the Joint Institute of Nuclear Research. PCT
thanks the Theory Group of Jefferson Lab for support and hospitality during a
sabbatical leave spent partly there in Fall 2005 and also thanks Craig Roberts for helpful
conversations.
\end{acknowledgments}

\bibliographystyle{apsrev}
\bibliography{refsPM,refsPCT,refsCDR,refs}

\end{document}